\documentstyle[aps,eqsecnum]{revtex}

\begin{document}

 \draft

\title{Effective field theory and the quark model, II. Structure of loop corrections}

\author{Loyal Durand\thanks{Electronic address: 
ldurand@theory2.physics.wisc.edu}, Phuoc Ha\thanks{Electronic address:
phuoc@theory1.physics.wisc.edu}, and Gregory Jaczko\thanks{Electronic
address:  Gregory.Jaczko@mail.house.gov}}
\address{ Department of Physics, University of Wisconsin-Madison,
Madison, WI 53706}

\date{\today}
\maketitle

\begin{abstract}
We analyze the structure of meson loop corrections to the O($m_s$\/) expressions for the baryon masses and magnetic moments in heavy-baryon chiral perturbation theory (HBChPT), and show in detail how the bulk of the corrections are absorbed into redefinitions of the unknown parameters of the O($m_s$\/) chiral expansion.  To effect this analysis, we use the three-flavor-index representation of the effective baryon fields and HBChPT developed in a preceding paper, and a decomposition of the corrections in terms of effective one-, two-, and three-quark operators. The results show why the loop corrections have so little effect on fits to the masses and moments, and do not seriously disrupt the Gell-Mann--Okubo relations for the masses and the Okubo relation for the moments even though individual loops can be quite large. We also examine the momentum structure of the residual loop corrections, and comment on limits on their validity in HBChPT. The structural analysis can be generalized straightforwardly to other problems in HBChPT using the three-flavor-index representation of the effective baryon fields, and provides a fairly simple way to determine what parts of the loop corrections are actually significant in a given setting.

\end{abstract}


\section{INTRODUCTION}
\label{sec:introduction}

Many authors have calculated meson loop corrections to the baryon masses \cite{Jen-HBChPT,Jen-masses,Bernard-masses,Bor-Mei-masses,Don-masses,%
DH-masses,phuoc-diss,jaczko-diss}, magnetic moments \cite{Caldi,Gasser,Krause,%
Jen-moments,Luty,Daietal,Mei-moments,DH-ChPTmoments,DH-loop-moments,%
Ha-moments}, weak currents \cite{Jen-HBChPT,Jen-axial}, and other quantities in various versions of chiral perturbation theory. The motivation is clear. Loop effects involving the relatively light pseudoscalar mesons should give the leading long-distance or low-momentum corrections to the short-distance chiral expansion, and can modify the structure of the theory. The results are striking. Individual loop corrections can be quite large, even so large in calculations using dimensional regularization that the convergence of the chiral expansion is questionable \cite{Don-masses,DH-ChPTmoments}. However, the effects on fits to the baryon masses and moments are generally small. The corrections do not, for example, lead to large violations of the Gell-Mann--Okubo relations for the masses \cite{DH-masses} or the Okubo relation for the moments
\cite{DH-ChPTmoments,DH-loop-moments}, both of which are satisfied by the short-distance expansions taken to O($m_s$\/) in the symmetry-breaking strange-quark mass, and to leading order in a momentum expansion.

The underlying reason for this result is that chiral perturbation theory is not a perturbation theory in the usual sense, with known expansion parameters. The short-distance parameters are not known {\em a priori}, but are rather determined in the fits to the data. All that is relevant in a final fit is the sum of the short-distance and loop contributions to each of the initial chiral invariants, and any new chiral invariants introduced by the loops \cite{DH-ChPTmoments}. It is only the latter that lead to violations of the O($m_s$\/) sum rules, and these new invariants are evidently small. This is not easy to see in the standard formulations given the complexity of the 
results.\footnote{See, for example, the expressions for the loop corrections to the masses and moments in \cite{DH-masses,DH-loop-moments}.} It is therefore be of interest to extract the new structures directly.

In the present paper, we give a detailed structural analysis of the loop corrections to the baryon masses, and a less detailed analysis for the moments. Our approach is based on the three-flavor-index representations of the effective octet and decuplet fields $B_{ijk}^{\,\gamma}$ and $T_{ijk}^{\,\mu\gamma}$ and the corresponding reformulation of heavy-baryon chiral perturbation theory (HBChPT) developed in a preceding paper \cite{DHJ}. This has the advantage of a close connection to quark-model ideas which can guide the decomposition of loop corrections in terms of effective one-, two-, and three-quark corrections. As we will show, the one-quark corrections can be absorbed completely in redefinitions of the chiral parameters, and substantial parts of the remaining corrections can be absorbed as well. The final results involve only effective operators which violate the O($m_s$\/) sum rules and can be extracted directly.

We also examine the momentum structure of the residual contributions to the masses. We find that, as usually calculated, they receive major contributions from momenta outside the the supposed range of validity of  ChPT. Rather mild structural assumptions consistent with dynamical models for the baryons eliminate this problem.

We believe that the methods developed here should be useful in other contexts as well, for example, in the analysis of axial currents and weak decays, and of low-energy meson-baryon scattering. They provide, in particular, a fairly transparent way of determining what quantities should be calculated in a given setting to get beyond the O($m_s$\/) chiral expansion.

The structure of the paper is as follows. In Sec.\ \ref{sec:background}, we review the properties of the effective baryon fields in the three-flavor-index notation we will use, write the Lagrangian of HBChPT in this notation, and introduce a compressed operator description of the meson-baryon couplings and the one- and two-body contributions to the baryon mass operator. The last corresponds exactly to the O($m_s$) mass operator in HBChPT, but is much simpler to analyze. In Sec.\ \ref{sec:loops-masses}, we analyze the structure of the one-loop corrections to the mass operator in detail, show that the bulk of the corrections can be absorbed in redefinitions of the parameters in the original mass operator, and isolate the new contributions that correct the Gell-Mann--Okubo mass formulas. In Sec.\ \ref{sec:moments}, we give the structure of the baryon magnetic-moment operator to O($m_s$), analyze the loop corrections to the leading terms, and show that the corrections can be absorbed completely in redefinitions of the input parameters. The only new structure, with associated changes to the Okubo relation for the moments, arises from loop corrections to terms in the moment operator that are already of O($m_s$). We summarize our conclusions in Sec.\ \ref{sec:conclusions}

\section{BACKGROUND: EFFECTIVE FIELD REPRESENTATION AND COUPLINGS}
\label{sec:background}

\subsection{Baryon and meson fields}
\label{subsec:fields}

We will work throughout the paper in terms of heavy-baryon chiral perturbation theory (HBChPT) \cite{Jen-HBChPT}. In this approach, the baryons in a process are supposed to be nearly on-shell with momenta $p^\mu=m_0v^\mu+k^\mu$, where $m_0$ is an appropriate baryon mass, $v^\mu$ is the four velocity of the on-shell baryon, $v_\mu v^\mu=1$, and $k^\mu$ is a small residual momentum. It is appropriate for such processes to rewrite the chiral Lagrangian by replacing the effective spin-1/2 octet and spin-3/2 decuplet baryon fields $B^{\gamma}$, $T^{\mu\gamma}$ by velocity-dependent fields $B^{\,\gamma}_v$, $T^{\,\mu\gamma}_v$ defined as \cite{georgi-HBPT}
\begin{eqnarray}
\label{B_v}
B_v(x) &=& \frac{1}{2}(1+\not \!v) e^{i m_0 {\not v} v^{\mu} x_{\mu}}B(x), \\
\label{T_v}
T^\mu_v(x) &=& \frac{1}{2}(1+\not \!v) e^{i m_0 {\not v} v^{\mu} x_{\mu}}T^\mu(x).
\end{eqnarray}
This transformation eliminates the large momentum $m_0v^\mu$ from the Dirac equation, and projects out particle rather than antiparticle operators. The velocity-dependent perturbation expansion of the redefined theory involves modified Feynman rules and an expansion in powers of $k/m_0$ \cite{georgi-HBPT}. The large mass $m_0$ does not appear directly in the new chiral expansion, the baryons may be treated as nonrelativistic in their quasi rest frame where $v^\mu=(1,\,\mbox{\boldmath$0$})$, and there are no baryon-antibaryon vertices at leading order in $k/m_0$. We will henceforth drop the velocity label on $B^{\,\gamma}_v$ and $T^{\,\mu\gamma}_v$, and deal only with the heavy baryon approximation at leading order.  

In writing the Lagrangian of HBChPT, we will follow the approach developed in detail in \cite{DHJ}, and will represent the octet and decuplet baryons by three-flavor-index spinor and vector-spinor fields $B_{ijk}^{\,\gamma}(x)$, $T_{ijk}^{\,\mu\gamma}$ which transform under matrices $U$ of the vector or diagonal subgroup SU(3)$_V$ of the chiral SU(3)$_L\times$SU(3)$_R$ as 
\begin{eqnarray}
\label{baryon_transform}
B_{ijk}^{\,\gamma} &\stackrel{U}{\longrightarrow}& U_{ii'}U_{jj'}U_{kk'} B_{i'j'k'}^{\,\gamma}, \nonumber \\
T_{ijk}^{\,\mu\gamma} &\stackrel{U}{\longrightarrow}& U_{ii'}U_{jj'}U_{kk'} T_{i'j'k'}^{\,\mu\gamma}.
\end{eqnarray}
Here $i,\,j,\,k\in u,\,d,\,s$ are flavor indices, $\gamma$ is a Dirac spinor index, and $\mu$ is a four-vector index. The fields are subject to the additional symmetry constraints
\begin{eqnarray}
\label{symmetries}
&& B_{jik}^{\,\gamma} = -B_{ijk}^{\,\gamma}, \\
\label{Jacobi}
&&B_{ijk}^{\,\gamma} + B_{jki}^{\,\gamma} + B_{kij}^{\,\gamma} = 0, \\
&&T_{ijk}^{\,\mu\gamma}=T_{jik}^{\,\mu\gamma}=T_{ikj}^{\,\mu\gamma}
\end{eqnarray}
necessary to get a flavor octet and decuplet. The free velocity-dependent fields also satisfy modified Dirac and Rarita-Schwinger constraints
\begin{eqnarray}
\label{Dirac_eqn_B}
\not\!v\,B_{ijk}&=&B_{ijk} ,\\
\not\!v\,T_{ijk}^{\,\mu} &=& T_{ijk}^{\,\mu}, \\
v_\mu T_{ijk}^{\,\mu} &=& 0.
\end{eqnarray}

We will use a normalization such that the $j_z=+1/2$ octet fields have the correspondence
\renewcommand{\arraystretch}{1.3}
\begin{equation}
\label{octet}
\begin{array}{c}
B_{121} \leftrightarrow \frac{1}{\!\!\sqrt{2}}\,p, \quad B_{122} \leftrightarrow \frac{1}{\!\!\sqrt{2}}\,n ,\\ 
B_{131} \leftrightarrow \frac{1}{\!\!\sqrt{2}}\,\Sigma^+, \quad B_{232} \leftrightarrow \frac{1}{\!\!\sqrt{2}}\,\Sigma^-, \\ 
B_{231} \leftrightarrow \frac{1}{\!\!\sqrt{2}}\,\Sigma^0 + \frac{1}{\!\!\sqrt{6}}\,\Lambda, \quad B_{132} \leftrightarrow \frac{1}{\!\!\sqrt{2}}\,\Sigma_0 -
\frac{1}{\!\!\sqrt{6}}\,\Lambda, \\
B_{133} \leftrightarrow \frac{1}{\!\!\sqrt{2}}\,\Xi^0, \quad B_{233} \leftrightarrow \frac{1}{\!\!\sqrt{2}}\,\Xi^- 
\end{array}
\end{equation}
\renewcommand{\arraystretch}{1}
with the physical particles, while the $j_z=+3/2$ decuplet fields have the correspondence
\renewcommand{\arraystretch}{1.3}
\begin{equation}
\label{decuplet}
\begin{array}{c}
T_{111} \leftrightarrow \Delta^{++},\quad T_{112} \leftrightarrow \frac{1}{\!\!\sqrt{3}}\,\Delta^+, \quad T_{122} \leftrightarrow \frac{1}{\!\!\sqrt{3}}\,\Delta^0,\quad T_{222} \leftrightarrow \Delta^-\\ T_{113} \leftrightarrow \frac{1}{\!\!\sqrt{3}}\,\Sigma^{*+}, \quad T_{123} \leftrightarrow \frac{1}{\!\!\sqrt{6}}\,\Sigma^{*0},\quad T_{223} \leftrightarrow \frac{1}{\!\!\sqrt{3}}\,\Sigma^{*-} \\ 
T_{133} \leftrightarrow \frac{1}{\!\!\sqrt{3}}\,\Xi^{*0},\quad T_{233} \leftrightarrow \frac{1}{\!\!\sqrt{3}}\,\Xi^{*-}, \quad T_{333} \leftrightarrow \Omega^-.
\end{array}
\end{equation}
\renewcommand{\arraystretch}{1}
The remaining $B$'s and $T$'s can be determined using the symmetry relations above. With these normalizations, we can sum over repeated indices in subsequent equations.

The $B$'s and $T$'s have the transformation properties of the spin-1/2 and spin-3/2 operators
\begin{eqnarray}
\label{Bijk}
B_{ijk}^{\,\gamma} &\longleftrightarrow& \frac{1}{6}\,\epsilon_{abc}\,q_i^{\,\alpha a}q_j^{\,\beta\, b}q_k^{\,\gamma\, c}\,(C\gamma^5)_{\alpha \beta}\,,\\
\label{Tijk}
T^{\,\mu\gamma}_{ijk}&\longleftrightarrow&\frac{1}{18\!\!\sqrt{2}}\epsilon_{abc}\left(q_i^{\,\alpha a} q_j^{\,\beta b} q_k^{\,\gamma c} + q_k^{\,\alpha a} q_j^{\,\beta b} q_i^{\,\gamma c} + q_i^{\,\alpha a} q_k^{\,\beta b} q_j^{\,\gamma c}\right) \left(C\gamma^\mu\right)_{\alpha\beta},
\end{eqnarray}
constructed from three anticommuting ``quark'' fields  $q_i^{\,\alpha a}$ which carry the color, flavor, and spin structure of the baryon.  Here $i$ and $\alpha$ are again flavor and spinor indices, while $a\in 1,\,2,\,3$ is a color index. Since we will not be dealing with color-dependent interactions, we will suppress the color indices and the completely antisymmetric factor $\epsilon_{abc}$, but will treat the $q$'s as commuting rather than anticommuting fields. We then have the correspondence
\begin{eqnarray}
\label{Bnocolor}
B_{ijk}^{\,\gamma} &\longleftrightarrow& \frac{1}{\sqrt6} \,\left(q_i^{\rm T} C\gamma^5q_j\right)q_k^{\,\gamma},\\
\label{Tnocolor}
T_{ijk}^{\,\mu\gamma}  &\longleftrightarrow& \frac{1}{6\sqrt{3}} \left[\,\left(q_i^{\rm T} C\gamma^\mu q_j\right) q_k^{\,\gamma} + j\leftrightarrow k + k\leftrightarrow i\,\right],
\end{eqnarray}
where the superscript T denotes a spinor transpose.

The effective octet pseudoscalar meson fields $\phi_{ij}$ correspond to quark-antiquark pairs in a singlet spin configuration,
\begin{equation}
\phi_{ij} \longleftrightarrow \frac{1}{\!\!\sqrt{6}}\,\left(q_i^{\,\alpha a}\bar{q}_j^{\, \beta b} - \frac{1}{3}\delta_{ij}\, q_k^{\,\alpha a}\bar{q}_k^{\, \beta b}\right)\,\delta_{ab}\,(C\gamma^5)_{\alpha \beta},
\end{equation}
and transform under SU(3)$_V$ as
\begin{equation}
\label{meson_trans}
\phi_{ij} \stackrel{U}{\longrightarrow} U_{ii'}U^*_{jj'}\phi_{i'j'} = U_{ii'}\phi_{i'j'}U^\dagger_{j'j}.
\end{equation}
The use of this representation makes the quark flow in an interaction diagram clear. However, physical meson masses are customarily used in loop calculations in chiral perturbation theory so we will use an alternative representation
in terms of the mass eigenstates  $\phi^l$, $l \in \pi,\,K,\,\eta$ in the following calculations, with 
\begin{equation}
\phi_{ij}= \frac{1}{2}\,\sum_{l=1}^8\,\lambda_{ij}^l\phi^{\,l}.
\end{equation}
Here the $\lambda$'s are appropriate combinations of the Gell-Mann matrices of SU(3). The inverse transformation gives the expected correspondence

\begin{equation}
\phi^{\,l} = \sum_{ij}\,\lambda_{ji}^l\phi_{ij} \longleftrightarrow \frac{1}{\!\!\sqrt{6}}\,\sum_{ij}\,\lambda_{ij}^lq_i^{\,\alpha a}\bar{q}_j^{\, \beta b}\,\delta_{ab}\,(C\gamma^5)_{\alpha \beta}.
\end{equation}
Note that we will not include the flavor-singlet pseudoscalar, nominally the $\eta'$, in our calculations.

The fields $B_{ijk}^{\,\gamma}$, $T_{ijk}^{\,\mu\gamma}$, and $\phi^l$ provide a complete set of effective fields for problems involving only the baryons and pseudoscalar mesons. As emphasized in \cite{DHJ}, the quark picture of these fields connects naturally to dynamical models, and may be used effectively to parametrize and analyze the most general spin-flavor structure of the matrix elements in HBChPT. However, the quarks in Eqs.\ (\ref{Bijk}) and (\ref{Tijk}) are not light dynamical quarks. They represent instead averages of the dynamical fields at leading order in the derivative expansion of HBChPT, carry no relative internal baryonic momenta, and therefore move with the baryons with the common four velocity $v^\mu$. Finally, as was illustrated in \cite{DHJ}, they may be treated as free quarks in determining the general spin-flavor structure of matrix elements. What are not determined  are the actual values of matrix elements. We will follow the same methods here, and treat the correspondences in Eqs.\ (\ref{Bijk}) and (\ref{Tijk}) as equalities. These expressions reduce in the rest frame of the baryon to the familiar nonrelativistic structures
\begin{eqnarray}
\label{BijkNR}
B_{ijk}^{\,\gamma} &=& -\frac{1}{\sqrt{6}}\left(q^{\rm T}_ii\sigma_2 q_j\right)q_k^{\,\gamma}, \\
{\bf T}_{ijk}^{\,\gamma} &=& \frac{1}{6\sqrt{3}} \left[\left(q_i^{\rm T} i\sigma_2\mbox{\boldmath$\sigma$}q_j\right)q_k^{\,\gamma} + j\leftrightarrow k + k\leftrightarrow i\right]\,,
\end{eqnarray}
where $T^\mu=(0,\,{\bf T})$ in the rest frame.

\subsection{The chiral Lagrangian}
\label{subsec:lagrangian}

We will write the Lagrangian of HBChPT in the form given in \cite{DHJ}. This is leading order in the derivative expansion and in the symmetry-breaking strange-quark mass parameter $m_s$. It can be constructed using standard methods by determining the most general chiral invariants with at most one derivative or factor of $m_s$. It can also be derived from the Lagrangian given, for example, in \cite{Jen-masses} by using the connection of the fields $B_{ijk}^{\,\gamma}$ with the standard matrix representation $B_{kl}^{\,\gamma}$ of the fields used there,
\begin{equation}
\label{Bkl}
B_{kl}^{\,\gamma} = \frac{1}{\!\!\sqrt{2}}\, \epsilon_{ijl}\,B_{ijk}^{\,\gamma}, \quad 
B_{ijk}^{\,\gamma} = \frac{1}{\!\!\sqrt{2}}\,\epsilon_{ijl}\,B_{kl}^{\,\gamma}.
\end{equation}
The $T$'s are the same. The result is
\begin{equation}
\label{L0+Lm}
{\cal L}={\cal L}_0+{\cal L}_M^B+{\cal L}_M^T,
\end{equation}
where the leading-order heavy-baryon Lagrangian ${\cal L}_0$ is
\begin{eqnarray}
\label{L0_B,T}
 {\cal L}_0 &=&  i\,\bar{B}_{kji}\left( v\!\cdot\!{\cal D}\, B\right)_{ijk}
+ 2(D+F)\, \bar{B}_{k'ji} A^\mu_{k'k} S_\mu B_{ijk} 
- 4(D-F)\, \bar{B}_{kji'} A^\mu _{i'i} S_\mu B_{ijk} \nonumber \\
&& -i\,\bar{T}_{kji}^\mu \left(v\!\cdot\!{\cal D}\, T_\mu\right)_{ijk}
+ 2\,{\cal H}\, \bar{T}_{kji'}^\mu A^\nu_{i'i} S_\nu T_{\mu;ijk}\\
&& +\sqrt{2}\,{\cal C} \left( \bar{T}_{kji'}^\mu A_{\mu;i'i} B_{ijk} + \bar{B}_{kji'} A_{\mu;i'i} T_{ijk}^\mu \right) + \frac{1}{4}f^2\,\left(\partial_\mu \Sigma_{ji}\,\partial^\mu\Sigma^\dagger_{ij}\right) \,, \nonumber
\end{eqnarray}
and the mass corrections ${\cal L}_M^B$ and ${\cal L}_M^T$ are
\begin{eqnarray}
\label{L^B_M}
{\cal L}^B_M &=& \delta m\,\bar{B}_{kji}\,B_{ijk} -\tilde{\alpha}_m\,\left(\bar{B}_{k'ji}^{\,\gamma}\,{\cal M}^+_{k'k}\,B_{ijk}^{\,\gamma} + \bar{B}_{kj'i}^{\,\gamma}\,{\cal M}^+_{j'j}\,B_{ijk}^{\,\gamma} + \bar{B}_{kji'}^{\,\gamma}\,{\cal M}^+_{i'i}\,B_{ijk}^{\,\gamma}\right) \nonumber \\
&&-2 \tilde{\alpha}_{ss} \left(4\bar{B}_{kji'}^{\,\gamma}\,{\cal M}^+_{i'i}\,B_{ijk}^{\,\gamma}-\bar{B}_{k'ji}^{\,\gamma}\,{\cal M}^+_{k'k}\,B_{ijk}^{\,\gamma}\right)  \\
\label{L^T_M}
{\cal L}^T_M &=& \delta m\,\bar{T}^\mu_{kji}\, T_{\mu;ijk} +  3(\tilde{\alpha}'_m-2\tilde{\alpha}'_{ss})\,\bar{T}_{k'ji}^{\,\mu\gamma}\,{\cal M}^+_{k'k}\,T_{\mu;ijk}^{\,\gamma}.
\end{eqnarray}

The bilinear chiral invariants above involve implied spinor products. $S^\mu$ is the spin operator defined in \cite{georgi-HBPT} and more formally in \cite{Jen-HBChPT}, 
\begin{equation}
S^\mu=\frac{1}{8}(1+\not\!v\,)\,\gamma^\mu\gamma^5(1+\not\!v\,).
\end{equation}
It reduces in the rest frame of the baryon to an ordinary spin operator acting on two-component rest spinors, $S^\mu\rightarrow (0,\,\mbox{\boldmath$\sigma$}/2)$. 

${\cal D}^\mu$ is the covariant derivative, 
\begin{eqnarray}
\label{covar_derivative}
{\cal D}^\nu B^{\,\gamma}_{ijk} &=& \partial^\nu B_{ijk}^{\,\gamma} + V_{ii'}^\nu B_{i'jk}^{\,\gamma} + V_{jj'}^\nu B_{ij'k}^{\,\gamma} + V_{kk'}^\nu B_{ijk'}^{\,\gamma},\\
{\cal D}^\nu T_{ijk}^{\,\mu\gamma} &=& \partial^\nu T_{ijk}^{\,\mu\gamma} + V_{ii'}^\nu T_{i'jk}^{\,\mu\gamma} + V_{jj'}^\nu T_{ij'k}^{\,\mu\gamma} + V_{kk'}^\nu T_{ijk'}^{\,\mu\gamma},
\end{eqnarray}
and $V_\mu$  and $A_\mu$ are the vector- and axial vector current matrices   
\begin{eqnarray}
\label{Vcurrent}
V_\mu &=&  \frac{1}{2}\left(\xi\,  \partial_\mu\xi^\dagger + \xi^\dagger\partial_\mu \xi\right) = (1/2f^2)\left[\,\phi\partial_\mu\phi - (\partial_\mu\phi) \phi\,\right]+\cdots \,,\\
\label{Acurrent}
A_\mu &=& \frac{i}{2}\left(\xi \, \partial_\mu \xi^\dagger - \xi^\dagger\partial_\mu \xi\right) =  f^{-1}\partial_\mu \phi+\cdots \,.
\end{eqnarray}
Here $\xi$ and $\Sigma$ are flavor matrices dependent on the meson fields $\phi$,
\begin{equation}
\label{xi}
\xi=e^{i\phi/f}, \quad \Sigma=e^{2i\phi/f}=\xi^{\,2},
\end{equation}
and $f \approx 93$ MeV is the pion decay constant. 

Finally, ${\cal M}^+$ is the flavor matrix
\begin{equation}
\label{Mpm}
{\cal M}_{\,ll'}^+ = \frac{1}{2}(\,\xi^\dagger M \xi^\dagger + \xi M\xi\,)_{ll'},
\end{equation}
where $M$ is the diagonal matrix 
\begin{equation}
\label{Mdefinition}
M = {\rm diag}\,(0,\,0,\,1\,).
\end{equation}

$D$, $F$, $\cal H$, and $\cal C$ in ${\cal L}_0$ are the couplings defined in \cite{Jen-masses}. In the mass terms, $\delta m=(m_\Delta-m_N)/2$, corresponding to our choice $m_0=(m_\Delta+m_N)/2$ for the reference mass extracted in the definition of the velocity-dependent fields in Eqs.\ (\ref{B_v}) and (\ref{T_v}). With this choice, all the remaining parameters are proportional to the strange quark mass $m_s$, a dependence we do not exhibit explicitly. They are related to the couplings $b_D$, $b_F$, and $c$ used in \cite{Jen-masses} by
\begin{equation}
\tilde{\alpha}_{ss}=\frac{2}{3}b_D m_s,\quad \tilde{\alpha}_m=-2b_F m_s -\frac{2}{3}b_D m_s,\quad \tilde{\alpha}'_m - 2\tilde{\alpha}'_{ss} = \frac{2}{3}c m_s.
\end{equation}

As discussed in \cite{DHJ}, the octet-decuplet mass shift $\delta m$ results from the spin dependence of the underlying interactions in QCD, and exists even in the symmetrical limit with $m_s=0$.\footnote{Note that we have included the $\delta m$ terms in the mass Lagrangians, and will treat the decuplet states explicitly in calculating loop corrections as in \cite{Jen-masses,Jen-moments}. This has been a subject of some disagreement in the literature, with, for example, the authors of \cite{Bor-Mei-masses} and \cite{Mei-moments} treating the decuplet states as heavy relative to the octet, and expanding their contribution to the chiral perturbation series in powers of $1/\delta m$. We do not find that approach convincing. The momenta and strange-meson masses that appear in loops are comparable to, or larger than $\delta m$, and convergence of the expansion is unlikely. In contrast, the perturbation expansion in $\delta m$ around  the symmetrical limit defined by ${\cal L}_0$ works well.} $\tilde{\alpha}_m$ is associated primarily with the effect of first-order strange-quark mass insertions on the octet masses, with some small extra contributions from the effect of spin-spin interactions on the dynamical matrix elements. $\tilde{\alpha}_{ss}$ is associated entirely with effective spin-spin interactions in the octet. $\tilde{\alpha}'_m$ and $\tilde{\alpha}'_{ss}$ have the same interpretation with respect to the decuplet masses. Since the last parameters only appear in the combination $\tilde{\alpha}'_m-2\tilde{\alpha}_{ss}$ in Eq.\ (\ref{L^T_M}), it requires dynamical information to separate them. As shown in \cite{DHJ}, the primed and unprimed parameters should, in fact, be approximately equal in the physically relevant case of weak spin-spin interactions.

\subsection{Meson-baryon couplings. Single-particle form}
\label{subsec:meson-baryon-couplings}

We expect the baryon-meson couplings in ${\cal L}_0$ to satisfy an approximate SU(6) symmetry for $m_s=0$ and small spin-spin interactions \cite{DHJ}. We will use this approximation, and write the couplings as
\begin{equation}
\label{SU6_couplings}
D=\beta, \quad F=\frac{2}{3}\beta, \quad {\cal C}=-2\beta, \quad {\cal H}=-3\beta,
\end{equation}
where $\beta$ is a common dynamical matrix element. The changes is these relations associated with quark masses and spin-dependent interactions only appear as higher order corrections in our later calculations.\footnote{The corrections to the effective axial couplings are similar to those for the baryon moment discussed in \cite{DHJ}, and are given for $m_s=0$ by the invariants (a), (d), and (j) in Eqs.\ (5.15) of that paper. The invariant (a) gives the SU(6) couplings above. The remaining invariants arise from spin-dependent interactions among the quarks, (d) from the analog of a spin-other-orbit coupling, and the two pieces of (j), from the different effects of short-range spin-spin interactions on the octet and decuplet wave functions. We expect the resulting changes to $D$, $F$, $\cal C$, and $\cal H$ to be small given the apparent weakness of the spin interactions in the baryons.} With this approximation, we can write all the interaction terms in Eq.\ (\ref{L0_B,T}) in terms of the single-particle operator
\begin{equation}
\label{axial_operator}
2\beta S_\mu A^\mu \;\raisebox{-1.5ex}{\mbox{$\stackrel{\textstyle\longrightarrow}{\scriptstyle{\bf v}\rightarrow 0}$}} \;\left(0,\,-\beta\mbox{\boldmath$\sigma$}\!\cdot\!{\bf A}\right)
\end{equation}
sandwiched between baryon operators, and taken to act on each quark in succession. For example, the octet-octet-meson terms in ${\cal L}_0$ are equivalent to
\begin{equation}
\label{L_BBM_rel}
{\cal L}_{BBM} = 2\beta\, \left[\bar{B}_{k'ji}\left(S^\mu\! A_\mu\right)_{k'k} B_{ijk} + \bar{B}_{kj'i}\left(S^\mu\! A_\mu \right)_{j'j}B_{ijk} + \bar{B}_{kji'}\left(S^\mu\! A_\mu\right)_{i'i}B_{ijk}\right],
\end{equation}
or, in the baryon rest frame, to
\begin{eqnarray}
\label{L_BBM-one-body}
{\cal L}_{BBM} &=& -\beta\, \left[\bar{B}_{k'ji}\,\left(\mbox{\boldmath$\sigma$}\! \cdot {\bf A}\right)_{k'k}\,B_{ijk} + \bar{B}_{kj'i}\,\left(\mbox{\boldmath$\sigma$}\! \cdot {\bf A}\right)_{j'j}\,B_{ijk} + \bar{B}_{kji'}\,\left(\mbox{\boldmath$\sigma$}\! \cdot {\bf A}\right)_{i'i}\,B_{ijk}\right] \nonumber \\
&=& -\beta\, \bar{B}_{k'j'i'}\left[\,\delta_{i'i}\delta_{j'j}\left(\mbox{\boldmath$\sigma$}\! \cdot {\bf A}\right)_{k'k} +\delta_{i'i}\left(\mbox{\boldmath$\sigma$}\! \cdot {\bf A}\right)_{j'j}\delta_{k'k} + \left(\mbox{\boldmath$\sigma$}\! \cdot {\bf A}\right)_{i'i}\delta_{j'j}\delta_{k'k}\,\right]B_{ijk}.  
\end{eqnarray}
The expressions for the octet-decuplet- and decuplet-decuplet-meson interactions are similar. The single-quark spin operator $\mbox{\boldmath$\sigma$}$ can be taken to act on either the final or initial quark, with its action defined by the replacement of $q$ by $\mbox{\boldmath$\sigma$}q$. A decomposition of the quark-level expressions in terms of the composite fields $B$ and $T$ reproduces the expression in Eq.\ (\ref{L0_B,T}) with the SU(6) couplings in Eq.\ (\ref{SU6_couplings}) as shown in \cite{DHJ}.

\begin{figure}
\centerline{
\epsffile{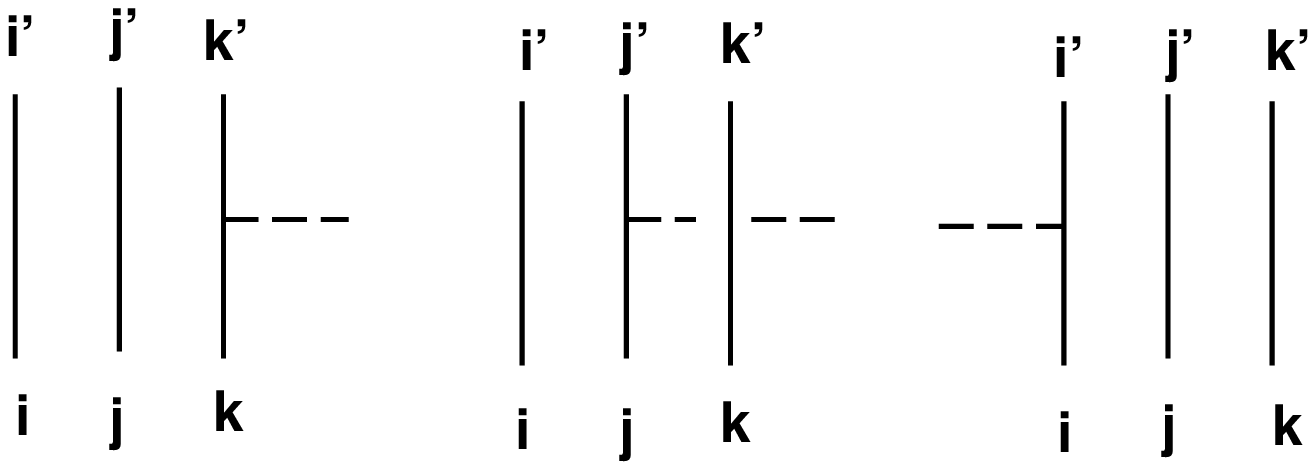}
}
\caption{Diagrammatic representation of the meson-baryon couplings at the quark level. Solid lines represent quark or flavor lines; dashed lines represent mesons. There is a factor $(-i\beta/2f)\,\mbox{\boldmath$\sigma$}\!\cdot\! {\bf k}\,\lambda^l_{p'p}$ at the vertex for quark $p$ interacting with meson $l$.}
\label{fig1}
\end{figure}

This description leads to an obvious diagrammatic representation of the interactions, with a baryon represented by a diagram with three quark lines. The the meson-baryon interaction then appears as a sum of meson-quark vertices distributed over the lines as in Fig.\ \ref{fig1}. The vertex connecting initial and final quarks $p,\,p'$ and an outgoing meson $l$  with four momentum $k^\mu$ is simply
\begin{equation}
\label{covar_meson_vertex}
\frac{2}{f}\beta\left<\,k\,|S^\mu\partial_\mu\phi_{p'p}|\,0\,\right> =\frac{i}{f}\beta\,S\!\cdot\!k\,\lambda^l_{p'p},
\end{equation}
where $\lambda$ is a Gell-Mann matrix in flavor space. There is an implied unit flavor matrix $\openone_{p'p}=\delta_{p'p}$ for any unmarked line. We will generally write the vertex factor in the rest frame of the baryon where it becomes 
\begin{equation}
\label{meson_vertex}
-\frac{i}{2f}\beta\,\mbox{\boldmath$\sigma$} \!\cdot\!{\bf k}\,\lambda^l_{p'p},
\end{equation}
and will work directly with the spin operators and $\lambda$ matrices in our analysis of loop corrections. The specialization to particular ingoing and outgoing hadrons will follow at the end when the final operators are sandwiched between the corresponding $B$'s and $T$'s.

\subsection{Baryon masses. One- and two-particle operators}
\label{subsec:mass_operators}

The mass terms in the baryon Lagrangian correspond to a Hamiltonian mass operator 
\begin{eqnarray}
\label{H_M}
{\cal H}_M &=& -\delta m\,\bar{B}_{kji}\,B_{ijk} -\delta m \,\bar{T}^\mu_{kji}\, T_{\mu;ijk} \nonumber \\
&&  +\tilde{\alpha}_m\,\left(\bar{B}_{k'ji}^{\,\gamma}\, {\cal M}^+_{k'k} \,B_{ijk}^{\,\gamma} + \bar{B}_{kj'i}^{\,\gamma}\,{\cal M}^+_{j'j} \,B_{ijk}^{\,\gamma} + \bar{B}_{kji'}^{\,\gamma}\,{\cal M}^+_{i'i} \,B_{ijk}^{\,\gamma}\right) -  3\tilde{\alpha}'_m \,\bar{T}_{k'ji}^{\,\mu\gamma} \,{\cal M}^+_{k'k} \,T_{\mu;ijk}^{\,\gamma}  \nonumber \\ 
&&+2 \tilde{\alpha}_{ss} \left(4\bar{B}_{kji'}^{\,\gamma}\,{\cal M}^+_{i'i} \,B_{ijk}^{\,\gamma}-\bar{B}_{k'ji}^{\,\gamma}\,{\cal M}^+_{k'k} \,B_{ijk}^{\,\gamma}\right) +6\tilde{\alpha}'_{ss} \,\bar{T}_{k'ji}^{\,\mu\gamma}\,{\cal M}^+ _{k'k}\,T_{\mu;ijk}^{\,\gamma}.
\end{eqnarray}
We can easily rewrite the baryon mass operators  in terms of one- and two-particle operators as was shown in \cite{DHJ}. We will start with the symmetrical limit in which $\tilde{\alpha}'_m =\tilde{\alpha}_m$ and $\tilde{\alpha}'_{ss} =\tilde{\alpha}_{ss}$, and will use ${\cal M}^+$ at leading order in the meson fields, ${\cal M}^+\rightarrow M$.

\begin{figure}
\centerline{
\epsffile{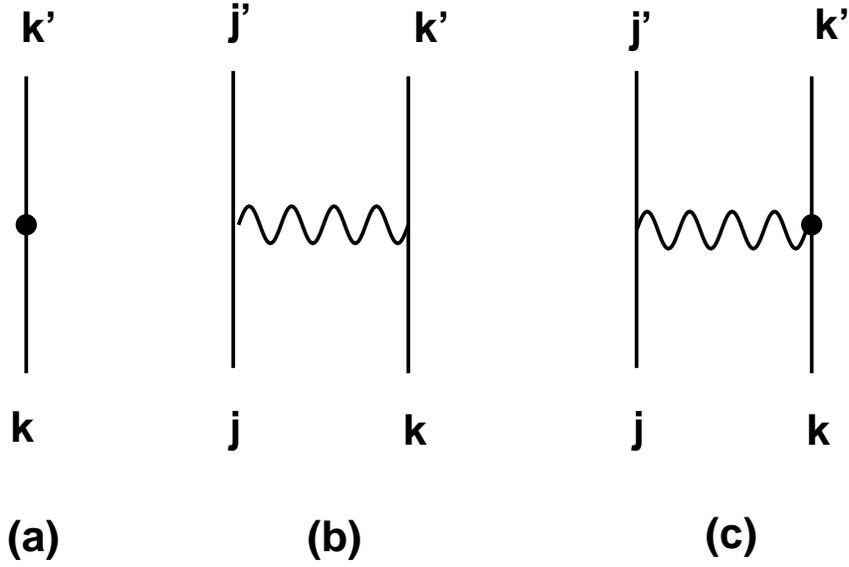}
}
\caption{Mass insertions in the symmetrical limit: (a) the single-particle strange-quark mass insertion with coefficient $\tilde{\alpha}_mM_{k'k}$;  (b)  the two-particle spin-spin mass insertion $\frac{1}{3}\delta m\, \mbox{\boldmath$\sigma$}_j\!\cdot\!\mbox{\boldmath$\sigma$}_k$ for the octet--decuplet mass; (c) the quark-mass-dependent spin-spin insertion with $-\tilde{\alpha}_{ss} \mbox{\boldmath$\sigma$}_j\!\cdot\, \mbox{\boldmath$\sigma$}_k M_{k'k}$\,. The wiggly lines in (b) and (c) denote a spin-spin interaction involving a dot product of Pauli matrices $\mbox{\boldmath$\sigma$}$ at the two vertices.  The dots in (a) and (c) denote the insertion of the flavor matrix $M$ at the corresponding vertex which necessarily lies on a strange-quark line.}
\label{fig2}
\end{figure}

The terms proportional to $\tilde{\alpha}_m$  in Eq.\ (\ref{H_M}) arise from  the single-particle operator
\begin{equation}
\label{m_s_insertion}
\tilde{\alpha}_m\left(\,\openone_{i'i}\openone_{j'j}M_{k'k} + \openone_{i'i}M_{j'j}\openone_{k'k} + M_{i'i}\openone_{j'j}\openone_{k'k}\,\right),
\end{equation}
where $\openone$ is a unit flavor matrix on a quark line. This operator is to be inserted, in the case of the octet, in a spinor product of the fields $\bar{B}_{k'j'i'}$, $B_{ijk}$. In the case of the decuplet, the operator must be inserted in a spinor--vector product of the physical rest-frame fields $\bar{\bf T}_{k'j'i'}$, ${\bf T}_{ijk}$. The result reproduces the $\tilde{\alpha}_m$ term in Eq.\ (\ref{H_M}) when we generalize it to an arbitrary Lorentz frame and use the flavor symmetry of the $T$'s. The operator above has an obvious diagrammatic interpretation, corresponding to the sucessive insertion of a ``quark-mass'' factor $\tilde{\alpha}_m M_{p'p}$ on each of the lines in a three-line baryon diagram. We will represent the insertion as in Fig.\ \ref{fig2}\,(a). 

The remaining terms in Eq.\ (\ref{H_M}) are associated with two-particle operators. Those proportional to $\delta m$ split the masses of the octet and decuplet states, and arise from the spin-spin operator 
\begin{equation}
\label{spin-spin_insertion}
\frac{1}{3}\delta m\left(\mbox{\boldmath$\sigma$}_i \!\cdot\! \mbox{\boldmath$\sigma$}_j + \mbox{\boldmath$\sigma$}_j\!\cdot\! \mbox{\boldmath$\sigma$}_k + \mbox{\boldmath$\sigma$}_k\!\cdot\! \mbox{\boldmath$\sigma$}_i\,\right)\openone_{i'i}\openone_{j'j}\openone_{k'k},
\end{equation}
where the index on a $\mbox{\boldmath$\sigma$}$ matrix indicates the quark on which the matrix acts. The complete spin operator has the values +3\,(-3) when acting on decuplet\,(octet) fields. However, in order to treat these fields symmetrically when the baryons appear as intermediate states in loops, we will retain the original expression, but associate each term with a two-particle diagram as shown in Fig.\ \ref{fig2}\,(b). The wiggly line in  the diagram indicates a spin-spin interaction with a factor $\mbox{\boldmath$\sigma$}$ at each quark vertex,  and an overall coefficient $\delta m/3$. We use the wiggly line to separate the vertices and display the spin and flavor structure of the interaction.  There is no associated propagator.

The terms in Eq.\ (\ref{H_M}) proportional to $\tilde{\alpha}_{ss}$ arise from the two-particle operator
\begin{equation}
\label{spin-spin_mass_insertion}
-\tilde{\alpha}_{ss}\left[\,(\mbox{\boldmath$\sigma$}_i+ \mbox{\boldmath$\sigma$}_j)
\!\cdot\! \mbox{\boldmath$\sigma$}_k\,\openone_{i'i}\openone_{j'j}M_{k'k} + (\mbox{\boldmath$\sigma$}_k+\mbox{\boldmath$\sigma$}_i)
\!\cdot\! \mbox{\boldmath$\sigma$}_j\,\openone_{i'i}M_{j'j}\openone_{k'k}
+ (\mbox{\boldmath$\sigma$}_j+\mbox{\boldmath$\sigma$}_k)
\!\cdot\! \mbox{\boldmath$\sigma$}_i\,M_{i'i}\openone_{j'j} \openone_{k'k} \right].
\end{equation}
This is a sum of spin-spin operators, but with one vertex distinguished by a factor of the strange-quark mass. We absorb this factor in $\tilde{\alpha}_{ss}$, but strange-quark lines are still picked out by the matrix $M={\rm diag}\,(0,\,0,\,1)$ with a unit entry in the $ss$ element. The presence of this matrix at a vertex is indicated by the dot in Fig.\ \ref{fig2}\,(c).

Combining the expressions above, we find that ${\cal H}_M$ is given by matrix elements of the spin-dependent operator
\begin{eqnarray}
\label{H_M_tree}
{\cal H}_{M,tree} &=& \frac{1}{3}\delta m\,
\left( \,\mbox{\boldmath$\sigma$}_i \!\cdot\! \mbox{\boldmath$\sigma$}_j + \mbox{\boldmath$\sigma$}_j\!\cdot\! \mbox{\boldmath$\sigma$}_k + \mbox{\boldmath$\sigma$}_k\!\cdot\! \mbox{\boldmath$\sigma$}_i\,\right) \nonumber \\
&& + \tilde{\alpha}_m\left(\,M_i + M_j + M_k\,\right) \\
&& - \tilde{\alpha}_{ss}\left[\,(\mbox{\boldmath$\sigma$}_i+ \mbox{\boldmath$\sigma$}_j)
\!\cdot\! \mbox{\boldmath$\sigma$}_k \,M_k + (\mbox{\boldmath$\sigma$}_k+\mbox{\boldmath$\sigma$}_i)
\!\cdot\! \mbox{\boldmath$\sigma$}_j\,M_j
+ (\mbox{\boldmath$\sigma$}_j+\mbox{\boldmath$\sigma$}_k)
\!\cdot\! \mbox{\boldmath$\sigma$}_i\,M_i\,\right] \nonumber
\end{eqnarray}
between the octet or decuplet fields. We have suppressed the unit flavor matrices or operators in Eqs.\ (\ref{m_s_insertion})--(\ref{spin-spin_mass_insertion}) for simplicity, and include only the nontrivial flavor operators $M$. The index on $M$ labels the quark line on which it acts. We will find it convenient later to work with this operator form of ${\cal H}_M$ rather than its explicit expression in terms of baryon fields.

We remark finally that the general mass operator ${\cal H}_M$ in Eq.\ (\ref{H_M}) with $\tilde{\alpha}_{m} \not= \tilde{\alpha}'_{m}$ and $\tilde{\alpha}_{ss} \not= \tilde{\alpha}'_{ss}$, as well as violations of the SU(6) symmetry of the meson-baryon couplings,  can be handled using the one- and two-body operators above with the new values of the octet and decuplet parameters, together with the projection operators for spins 1/2 and 3/2,
\begin{eqnarray}
\label{projection_ops}
P_{1/2} &=& \frac{1}{6}\left(\,3 - \mbox{\boldmath$\sigma$}_i \!\cdot\! \mbox{\boldmath$\sigma$}_j - \mbox{\boldmath$\sigma$}_j\!\cdot\! \mbox{\boldmath$\sigma$}_k - \mbox{\boldmath$\sigma$}_k\!\cdot\! \mbox{\boldmath$\sigma$}_i\,\right), \\
P_{3/2} &=&\frac{1}{6}\left(\,3 + \mbox{\boldmath$\sigma$}_i \!\cdot\! \mbox{\boldmath$\sigma$}_j + \mbox{\boldmath$\sigma$}_j\!\cdot\! \mbox{\boldmath$\sigma$}_k + \mbox{\boldmath$\sigma$}_k\!\cdot\! \mbox{\boldmath$\sigma$}_i\,\right).
\end{eqnarray}
As discussed in \cite{DHJ}, the differences between the primed and unprimed parameters result from the effects of the weak spin-dependent interactions between quarks, and involve the appearance of three-body operators of the form $\mbox{\boldmath$\sigma$}_i\!\cdot\! \mbox{\boldmath$\sigma$}_j M_{k'k}$ in the effective field theory.  The differences in the parameters are relatively small in fits to the masses. We will therefore ignore this refinement---a correction to a correction--- in discussing meson loop corrections, though not in the mass terms given by ${\cal H}_M$.

\section{ONE-LOOP CORRECTIONS TO BARYON MASSES}
\label{sec:loops-masses}

\subsection{Limit of degenerate baryon masses}
\label{subsec:degenerate-masses}

\begin{figure}
\centerline{
\epsffile{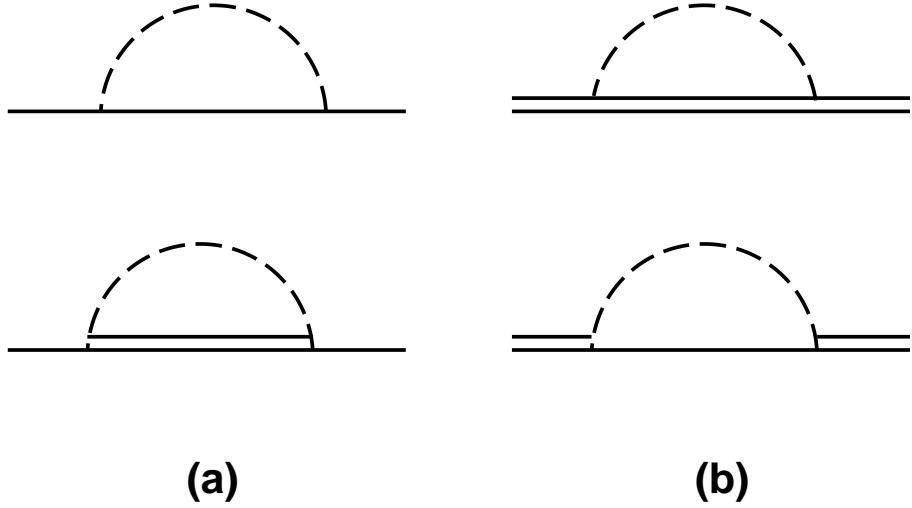}
}
\caption{The one-loop contributions to (a), the octet mass, and (b), the decuplet mass, as seen at the at the baryon level. Single and double lines represent octet and decuplet states, respectively.}
\label{fig3}
\end{figure}

We will begin our analysis of loop corrections to the baryon masses starting from the limit of degenerate masses within and between the octet and decuplet multiplets as in \cite{Jen-masses}, and will consider the effect of mass insertions in the following section. We show the baryon-level graphs corresponding to the one-loop mass corrections in Fig.\ \ref{fig3}\,(a) for the octet baryons, and in Fig.\ \ref{fig3}\,(b) for the decuplet. There are octet and decuplet intermediate states in each case. These are not distinguished by the particle masses, and the couplings at the vertices are related by the SU(6) relations discussed above. 

\begin{figure}
\centerline{
\epsffile{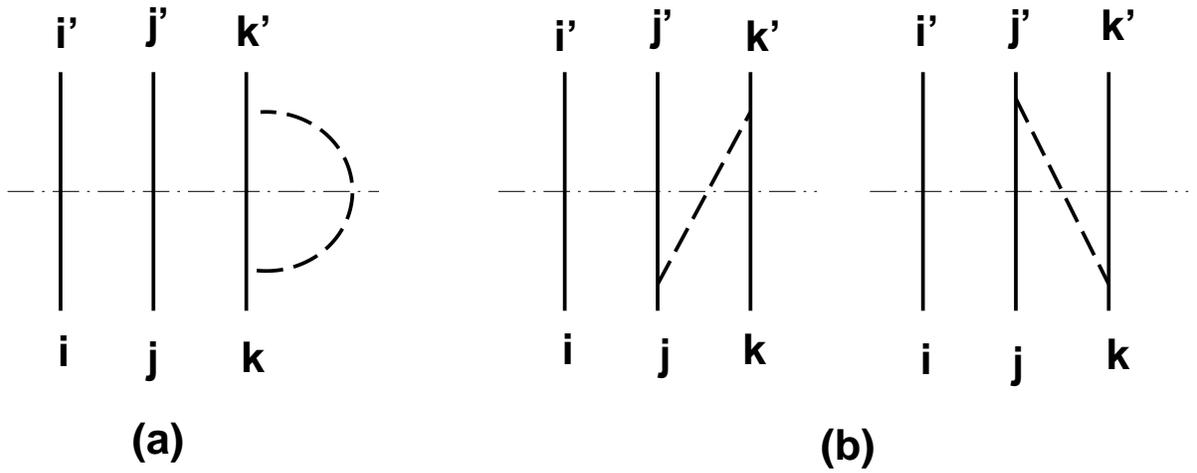}
}
\caption{Components of the one-loop baryon self-energy graphs of Fig.\ {\protect\ref{fig3}} seen as time-ordered graphs at the quark level. Solid lines represent quarks. The horizontal dot-dashed line picks out the intermediate state. (a): a self-energy graph. (b): exchange graphs. The associated renormalization graphs have the same structure, but involve different momentum integrals $I_l$ and $I'_l$. The baryon-level graphs in Fig.\ {\protect\ref{fig3}} are obtained by projecting the operator structure of the quark graphs on specific external and intermediate baryon states.}
\label{fig4}
\end{figure}

We can decompose the baryon graphs in Fig.\ \ref{fig3} at the quark level using the expression for the coupling operator in Eq.\ (\ref{meson_vertex}). The components divide into self-energy and exchange graphs as shown in Figs.\ \ref{fig4}\,(a) and (b). We use time-ordered perturbation diagrams in this figure to display the loop structure and the intermediate states. The graphs in Fig.\ \ref{fig3} are obtained by projecting the operator structure of the quark graphs on specific external and intermediate baryon states. In particular, the octet and decuplet states are complete at leading order in the derivative expansion of ChPT, with the effects of baryon structure and excited states appearing only at higher orders.

Note that the quarks do not carry relative momenta within the baryon in the heavy-baryon limit \cite{DHJ}, and in that sense are not dynamical. They move with the heavy baryon, and the energy denominators corresponding to the intermediate states in Figs.\ \ref{fig3} and \ref{fig4} are all equal to $1/[E_0-E({\bf k})]=-1/\sqrt{{\bf k}^2+M_l^2}$, where $M_l$ is the mass of the meson in the loop and $\bf k$ is its momentum in the baryon rest frame. The loop integrals therefore all have the same momentum structure, though with different meson masses. In particular, the diagrams in Fig.\ \ref{fig4}\,(b) give contributions 
\begin{equation}
\label{fig4b}
\frac{\beta^2}{4f^2}\sum_l\int\frac{d^3k}{(2\pi)^3\,2\sqrt{{\bf k}^2+M_l^2}}\lambda^l_{j'j}\,\mbox{\boldmath$\sigma$}_j\!\cdot\!{\bf k}\,\frac{1}{E_0-E({\bf k})}\,\mbox{\boldmath$\sigma$}_k\!\cdot\!{\bf k}\,\lambda^l_{k'k}\,F^2({\bf k^2}) = -\frac{1}{3}\mbox{\boldmath$\sigma$}_j \!\cdot\! \mbox{\boldmath$\sigma$}_k\sum_l\lambda^l_{j'j}\lambda^l_{k'k} I_l
\end{equation}
to the mass operator. Here $I_l$ is the common integral for the exchange of meson $l$,\footnote{The same integral is obtained in Feynman perturbation theory with Pauli-Villars regularization when the four-dimensional integration is performed by integrating first over the $k^0$ component of $k^\mu$, and the result is evaluated in the rest frame of the baryon.} 
\begin{equation}
\label{I_l}
I_l = \frac{\beta^2}{16\pi^2f^2}\int_0^\infty dk\,  \frac{k^4}{k^2+M_l^2}\,F^2(k^2),
\end{equation}
and $F(k^2)$ is a form factor which limits the integration to low momenta. The contribution of the diagram in Fig.\ \ref{fig4}\,(a) is given similarly by
\begin{equation}
\label{fig4a}
-\frac{1}{3}\, \mbox{\boldmath$\sigma$}_k \!\cdot\! \mbox{\boldmath$\sigma$}_k\,\sum_l\lambda^l_{k'k''}\lambda^l_{k''k}I_l = -\sum_l\left(\lambda^l\lambda^l\right)_{k'k}\,I_l.
\end{equation}

The cutoff factor $F$ arises physically from the composite structure of the baryons. In particular, a baryon will not remain in its ground state when a dynamical quark is given a high recoil momentum relative to the meson, and the matrix element is correspondingly damped through the extended structure of the baryon and meson wave functions. It is then not legitimate to work at leading order in the momentum expansion. $F^2$ may be regarded alternatively as a Pauli-Villars convergence factor introduced to make the theory finite, and to restrict the loop integrations to momenta well below the chiral cutoff at which an effective field theory written in terms of $B$, $T$, and $\phi$ fails.

The topology of the ``self-energy diagram'' in Fig.\ \ref{fig4}\,(a) suggests that this contribution is equivalent to a shift in the effective quark masses. We will show that it leads, in fact, to a shift in the mass parameters $m_0$ and $\tilde{\alpha}_m$, and has no other effect. The spin-spin structure of the exchange terms suggests also, as we show below, that parts of those terms can be absorbed by redefinitions of the parameters $\delta m$ and $\tilde{\alpha}_{ss}$. The only new structure introduced by these diagrams results from the appearance of $\lambda$ matrices on two different lines in the exchange diagrams. To demonstrate these results, we need to express the products of $\lambda$ matrices in Eqs.\ (\ref{fig4b}) and (\ref{fig4a}) in terms of the matrices $\openone$ and $M$ which appear in Eqs.\ (\ref{m_s_insertion})--(\ref{spin-spin_mass_insertion}), plus the only new structure that can appear for two-quark correlations, the product of two $M$'s with one on each line.

The product in Eq.\ (\ref{fig4a}) can be reduced by using the commutation and anticommutation relations of SU(3), or by direct calculation, with the result
\begin{equation}
\label{lambda_lambda_sum}
\sum_lI_l\left(\lambda^l\lambda^l\right)_{k'k} = \left(3I_\pi+2I_K+\frac{1}{3}I_\eta\right)\openone_{k'k} - \left(3I_\pi-2I_K -I_\eta\right)\,M_{k'k}. 
\end{equation}
Here $M$ is the matrix in Eq.\ (\ref{Mdefinition}), $I_l=I_\pi$ for $i=1$--3, $I_l=I_K$ for $l=4$--7, and $I_l=I_\eta$ for $l=8$. The term proportional to $M$ would vanish for equal meson masses. It is nonzero only because of the effect of insertions of $m_s$ on those masses. The total contribution of the three quark self-energy diagrams gives an effective operator
\begin{equation}
\label{delta_H_se}
\delta{\cal H}_{M,se} = -\left(9I_\pi+6I_K +I_\eta\right)  + \left(3I_\pi-2I_K -I_\eta\right)\left(M_i + M_j + M_k\right),
\end{equation}
where we again display only the nontrivial flavor operators.

The time-ordered exchange graphs in Fig.\ \ref{fig4}\,(b) each give a contribution
\begin{equation}
\label{exchange_graph}
-\frac{1}{3}\sum_l\lambda^l_{j'j}\lambda^l_{k'k}\,I_l\,\mbox{\boldmath$\sigma$}_j\!\cdot\!\mbox{\boldmath$\sigma$}_k
\end{equation}
to the baryon self energy. The $\lambda$ matrices now appear in a four-index flavor tensor with a somewhat complicated structure. This may be reduced by writing out the product of $\lambda$ matrices explicitly in terms of their entries, rearranging terms guided by the possible flavor flow through the exchanged meson, and introducing factors of the matrix $M$ to pick out strange-quark indices. For example, 
\begin{equation}
\label{lambda_lambda_pi}
\sum_{l=1}^3\lambda^l_{j'j}\lambda^l_{k'k}\,I_l = \left(2\delta_{j'k}\delta_{k'j} - \delta_{j'j}\delta_{k'k}\right)\,I_\pi,
\end{equation}
where $j,\,k\in 1,\,2$ only. We can extend this expression to include the third flavor by the substitutions
\begin{equation}
\delta_{j'k}\longrightarrow \left(\openone-M\right)_{j'k}, \quad \delta_{j'j}\longrightarrow \left(\openone-M\right)_{j'j}.
\end{equation}
Using similar constructions for $l=3$--8, combining terms, and noting that $M_{j'k}M_{k'j}=M_{j'j}M_{k'k}$,  we find that 
\begin{eqnarray}
\label{I_lambda'_lambda}
\sum_l\lambda^l_{j'j}\lambda^l_{k'k}\,I_l &=& 2I_\pi\,\openone_{j'k}\openone_{k'j}  - \left(I_\pi-\frac{1}{3}I_\eta\right)\,\openone_{j'j}\openone_{k'k}  \nonumber \\
&& - 2\left(I_\pi-I_K\right) \left(\openone_{j'k}M_{k'j} + M_{j'k}\openone_{k'j}\right)   \nonumber \\ 
&& +\left(I_\pi-I_\eta\right) \left(\openone_{j'j}M_{k'k} + M_{j'j}\openone_{k'k}\right) \\
&& +\left(I_\pi-4I_K+3I_\eta\right)M_{j'j}M_{k'k}  \nonumber
\end{eqnarray}
%
%\, \mbox{\boldmath$\sigma$}_j\!\cdot\!\mbox{\boldmath$\sigma$}_k 
%\label{exchange_graphs}

The first and third terms in this equation involve structures that do not appear in Eq.\ (\ref{H_M_tree}). However, when multiplied by $\mbox{\boldmath$\sigma$}_j\!\cdot\!\mbox{\boldmath$\sigma$}_k$ as in Eq.\ (\ref{exchange_graph}), these terms can be reduced to the existing structures by using the spin- or flavor-exchange operator \cite{DHJ} 
\begin{equation}
\label{P_2}
P_{j,k} = \frac{1}{2}\left(\,1 + \mbox{\boldmath$\sigma$}_j\!\cdot\!\mbox{\boldmath$\sigma$}_k \, \right),
\end{equation}
and the properties of the Pauli matrices. Thus, recalling that $j,\,k$ label the quark lines on which those matrices act, 
\begin{eqnarray}
\openone_{j'k}\openone_{k'j}\,\mbox{\boldmath$\sigma$}_j\!\cdot\!
\mbox{\boldmath$\sigma$}_k  &=& P_{j'k'}\openone_{j'j}\openone_{k'k}\,\mbox{\boldmath$\sigma$}_j\!\cdot\!
\mbox{\boldmath$\sigma$}_k \nonumber \\
&=& \frac{1}{2}\openone_{j'j}\openone_{k'k}\left(\,1 + \mbox{\boldmath$\sigma$}_j\!\cdot\!\mbox{\boldmath$\sigma$}_k\,\right) \mbox{\boldmath$\sigma$}_j\!\cdot\!\mbox{\boldmath$\sigma$}_k \\
&=& \left(\frac{3}{2} - \frac{1}{2}\, \mbox{\boldmath$\sigma$}_j\!\cdot\!\mbox{\boldmath$\sigma$}_k\right) \openone_{j'j}\openone_{k'k}. \nonumber
\end{eqnarray}
Similarly,
\begin{equation}
\left(\openone_{j'k}M_{k'j} + M_{j'k}\openone_{k'j}\right) \, \mbox{\boldmath$\sigma$}_j\!\cdot\!\mbox{\boldmath$\sigma$}_k = \left(\frac{3}{2} - \frac{1}{2}\, \mbox{\boldmath$\sigma$}_j\!\cdot\!\mbox{\boldmath$\sigma$}_k\right) \left(\openone_{j'j}M_{k'k} + M_{j'j}\openone_{k'k}\right).
\end{equation}
Using these identities, we find that 
\begin{eqnarray}
\label{exchange_graphs}
\sum_lI_l\,\lambda^l_{j'j}\lambda^l_{k'k}\, \mbox{\boldmath$\sigma$}_j\!\cdot\!\mbox{\boldmath$\sigma$}_k &=& 3I_\pi\,\openone_{j'j}\openone_{k'k}  - \left(2I_\pi-\frac{1}{3}I_\eta\right) \openone_{j'j}\openone_{k'k}\,\mbox{\boldmath$\sigma$}_j\!\cdot\!\mbox{\boldmath$\sigma$}_k  \nonumber \\
&& -3\left(I_\pi-I_K\right)\left(\openone_{j'j}M_{k'k}+M_{j'j} \openone_{k'k}\right) \nonumber \\
&& + \left(2I_\pi-I_K-I_\eta\right) \left(\openone_{j'j}M_{k'k'}  + M_{j'j}\openone_{k'k}\right)\, \mbox{\boldmath$\sigma$}_j\!\cdot\!\mbox{\boldmath$\sigma$}_k   \nonumber \\ 
&& +\left(I_\pi-4I_K+3I_\eta\right) M_{j'j}M_{k'k}\, \mbox{\boldmath$\sigma$}_j\!\cdot\!\mbox{\boldmath$\sigma$}_k  
\end{eqnarray}

Using this relation in Eq.\ (\ref{exchange_graph}), adding in similar contributions from the $i,\,j$ and $k,\,i$ exchange graphs not shown in Fig.\ \ref{fig4}\,(b), and combining the results with the self-energy contributions in Eq.\ (\ref{delta_H_se}), we obtain a one-loop change in the mass operator given by
\begin{eqnarray}
\label{delta_H_M_loop}
\delta{\cal H}_{M,loop} &=& -\left(15I_\pi+6I_K+I_\eta\right) +\frac{2}{3}\left(2I_\pi-\frac{1}{3}I_\eta\right) \left(\mbox{\boldmath$\sigma$}_{i} \!\cdot\! \mbox{\boldmath$\sigma$}_{j} + \mbox{\boldmath$\sigma$}_{j} \!\cdot\! \mbox{\boldmath$\sigma$}_{k} + \mbox{\boldmath$\sigma$}_{k} \!\cdot\! \mbox{\boldmath$\sigma$}_{i}\right) \nonumber \\
&& + \left(I_\pi-2I_K+I_\eta\right)\left(\,M_i + M_j + M_k\, \right)\nonumber \\
&& -\frac{2}{3}\left(2I_\pi-I_K-I_\eta\right)\left[\,M_i\mbox{\boldmath$\sigma$}_{i} \!\cdot\! \left(\mbox{\boldmath$\sigma$}_{j} + \mbox{\boldmath$\sigma$}_{k}\right) + M_j\mbox{\boldmath$\sigma$}_{j} \!\cdot\! \left(\mbox{\boldmath$\sigma$}_{k} + \mbox{\boldmath$\sigma$}_{i}\right) M_k\mbox{\boldmath$\sigma$}_{k} \!\cdot\! \left(\mbox{\boldmath$\sigma$}_{i} + \mbox{\boldmath$\sigma$}_{j}\right)\, \right]  \nonumber\\
&& -\frac{2}{3}\left(I_\pi-4I_K+3I_\eta\right)\left[\, M_iM_j \mbox{\boldmath$\sigma$}_{i} \!\cdot\! \mbox{\boldmath$\sigma$}_{j} + M_jM_k \mbox{\boldmath$\sigma$}_{j} \!\cdot\! \mbox{\boldmath$\sigma$}_{k} + M_kM_i \mbox{\boldmath$\sigma$}_{k} \!\cdot\! \mbox{\boldmath$\sigma$}_{i}\right],
\end{eqnarray}
where we continue to suppress unit flavor operators for simplicity. 

The first term in $\delta{\cal H}_{M,loop}$ shifts the overall mass $m_0$ to $m_0-(15I_\pi+6I_K+I_\eta)$, and can be absorbed in a redefinition of $m_0$. The second, third, and fourth terms have the same structure as the corresponding terms in ${\cal H}_{M,tree}$, Eq.\ (\ref{H_M_tree}). Thus, as expected, there is no change in the overall structure of the mass operator through first order in $M$. All the possible chiral invariants to this order were already included in ${\cal H}_{M,tree}$, and corresponding terms in ${\cal H}_{M,tree}$ and $\delta{\cal H}_{M,loop}$ can be combined. The last term in Eq.\ (\ref{delta_H_M_loop}) involves $MM$, with the operators on different lines,  and is new. This term vanishes for equal meson masses and has the coefficient structure of the Gell-Mann--Okubo mass relation for mesons, properties which will be important later, and is formally of order $m_s^{3/2}$.

\subsection{Renormalization constants and mass insertions}
\label{subsec:mass-insertions}

In dealing below with the effects of baryon mass splittings on the one-loop self energies, we will need the wave function renomalization constants $Z=1-\delta Z$. The one-loop contributions to $\delta Z$ have the same spin and flavor structure as the mass diagrams in Fig.\ \ref{fig4}, so have components with the self-energy structure for each of the quark lines, plus components corresponding to each of the six time-ordered exchange graphs. However, the momentum integrals have an extra energy denominator\footnote{
Specifically, $Z=1-\sum_n\frac{\left<0|H'|n\right>\left<n|H'|0\right>}{(E_0-E_n)^2}$, while $E^{(2)}=\sum_n\frac{\left<0|H'|n\right>\left<n|H'|0\right>} {E_0-E_n}$.}
 so the common baryon-level integral $I_l$ is replaced by a modified integral $I'_l$ given by 
\begin{equation}
\label{I'_l}
I'_l = \frac{\beta^2}{16\pi^2f^2}\int_0^\infty dk\,  \frac{k^4}{(k^2+M_l^2)^{3/2}}\,
F^2(k^2).
\end{equation}
With this definition,
\begin{eqnarray}
\label{deltaZ}
\delta Z_{i'j'k';ijk} &=& \sum_l I'_l\left[\,\openone_{i'i}\openone_{j'j} \left(\lambda^l\lambda^l\right)_{k'k} + \openone_{i'i} \left(\lambda^l\lambda^l\right)_{j'j}\openone_{k'k}  + \left(\lambda^l\lambda^l\right)_{i'i}\openone_{j'j}\openone_{k'k}\,  \right] \nonumber \\
&& +\frac{1}{3}\sum_l I'_l\left(\,\openone_{i'i} \lambda^l_{j'j} \lambda^l_{k'k} \,\mbox{\boldmath$\sigma$}_j\!\cdot\!\mbox{\boldmath$\sigma$}_k + \lambda^l_{i'i} \openone_{j'j} \lambda^l_{k'k} \,\mbox{\boldmath$\sigma$}_i\!\cdot\!\mbox{\boldmath$\sigma$}_k + \lambda^l_{i'i} \lambda^l_{j'j} \openone_{k'k} \,\mbox{\boldmath$\sigma$}_i\!\cdot\!\mbox{\boldmath$\sigma$}_j\,\right),
\end{eqnarray}

The changes in the loop contributions which arise from the baryon mass splittings have a more complicated operator structure. We defined the basic loop integral $I_l$ in Eq.\ (\ref{fig4b}) for degenerate baryon masses $m_0$. This mass was removed in the transformation to the heavy-baryon fields and does not appear in the energy denominator. If we include the baryon mass differences generated by ${\cal H}_{M,tree}$, $E({\bf k})$ is replaced in the energy denominator by  $E({\bf k})+{\cal H}_{M,tree}$. Expanding to first order, we obtain extra contributions to the expressions for the integrals in Eqs.\ (\ref{fig4b}) and (\ref{fig4a}) given schematically by
\begin{eqnarray}
\label{fig4b'}
\frac{\beta^2}{4f^2}\sum_l\int\frac{d^3k}{(2\pi)^3\,2\sqrt{{\bf k}^2+M_l^2}}\frac{F^2({\bf k}^2)}{[E_0-E({\bf k})]^2}\,&&\left(\lambda^l \,\mbox{\boldmath$\sigma$}_j\!\cdot\!{\bf k}  \, {\cal H}_{M,tree} \,\mbox{\boldmath$\sigma$}_k \!\cdot\! {\bf k}\,\lambda^l\right)_{j'k';jk} \nonumber \\
&& = \frac{1}{3}\sum_l I'_l\, \left(\lambda^l\, \mbox{\boldmath$\sigma$}_j \!\cdot\!{\cal H}_{M,tree} \mbox{\boldmath$\sigma$}_k\,\lambda^l\right)_{j'k';jk}.
\end{eqnarray}
In the case of the self-energy type diagrams, Fig. \ref{fig4}\,(a), the $\lambda$ matrices and $\mbox{\boldmath$\sigma$}$'s appear at vertices on the same quark line, with one preceding and the other following the mass insertion. In the case of the exchange diagrams, Fig.\ \ref{fig4}\,(b), the vertices are on different lines and appear with two time orderings. 

\begin{figure}
\centerline{
\epsffile{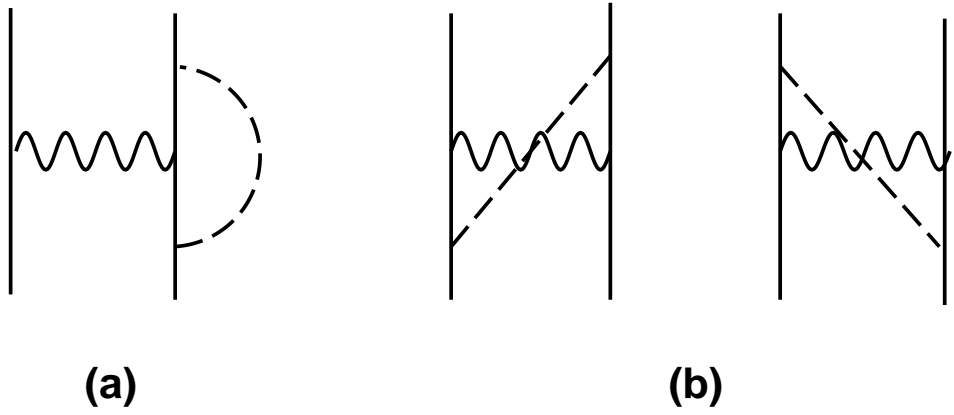}
}
\caption{One-loop self-energy and exchange graphs which contribute to the octet-decuplet mass splitting, but not to splitting within the multiplets.}
\label{fig5}
\end{figure}

\begin{figure}
\centerline{
\epsffile{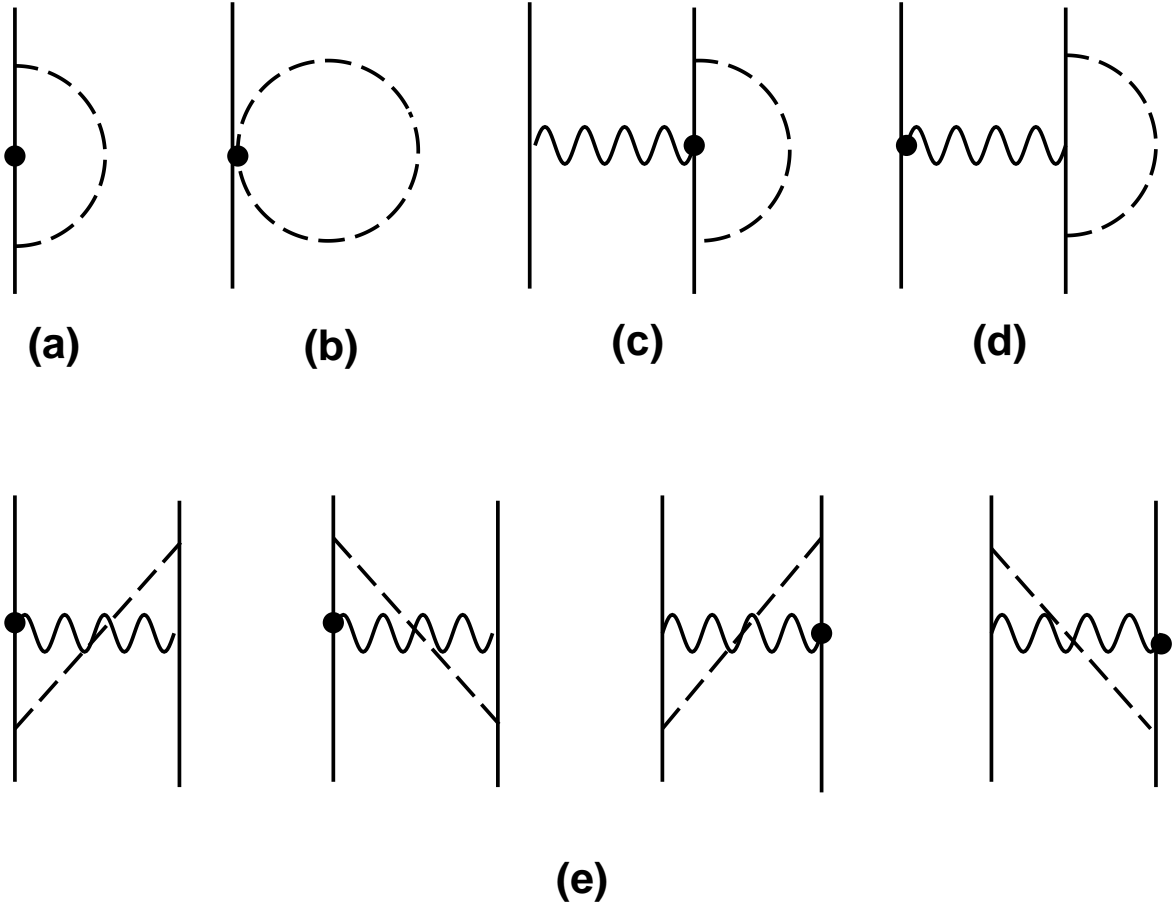}
}
\caption{$M$-dependent mass operators at one loop. (a): one-body diagram involving the insertion of the quark mass operator in Fig.\ {\protect\ref{fig2}}\,(a) in the one-body diagram in  Fig.\ {\protect\ref{fig4}}\,(a). (b): O($\phi^2$) closed-loop correction to Fig.\ {\protect\ref{fig4}}\,(a) from the expansion of ${\cal M}^+$. (c)--(e): two-body operators involving insertions of the $M$-dependent two-body operator in Fig.\ {\protect\ref{fig2}}\,(c) in the self-energy and exchange diagrams of Fig.\ {\protect\ref{fig4}}, shown as time-ordered diagrams. The wiggly line is a point spin-spin interaction with no propagator or time ordering. Dots indicate insertions of the flavor matrix $M$.  }
\label{fig6}
\end{figure}

The corrections for baryon mass insertions are described diagrammatically in Figs.\ \ref{fig5} and \ref{fig6}. With the exception of the diagram in Fig.\ \ref{fig6}\,(b) which we will discuss later, the corrections are given analytically by $I'_l$ multiplied by an operator coefficient determined by the relevant diagram. The full effect of the mass insertions also includes the renormalization of the basic mass operator ${\cal H}_{M,tree}$ by initial and final factors $Z^{1/2}$, giving a one-loop renormalization correction  
\begin{equation}
\label{mass_renorm}
-\frac{1}{2}\left(\delta Z\,{\cal H}_{M,tree} + {\cal H}_{M,tree} \delta Z\right)_{j'k';jk}.
\end{equation}

\begin{figure}
\def\picsize{6in}                  
\def\picfilename{fig7.eps}         
\centerline{\epsfxsize\picsize\epsfbox{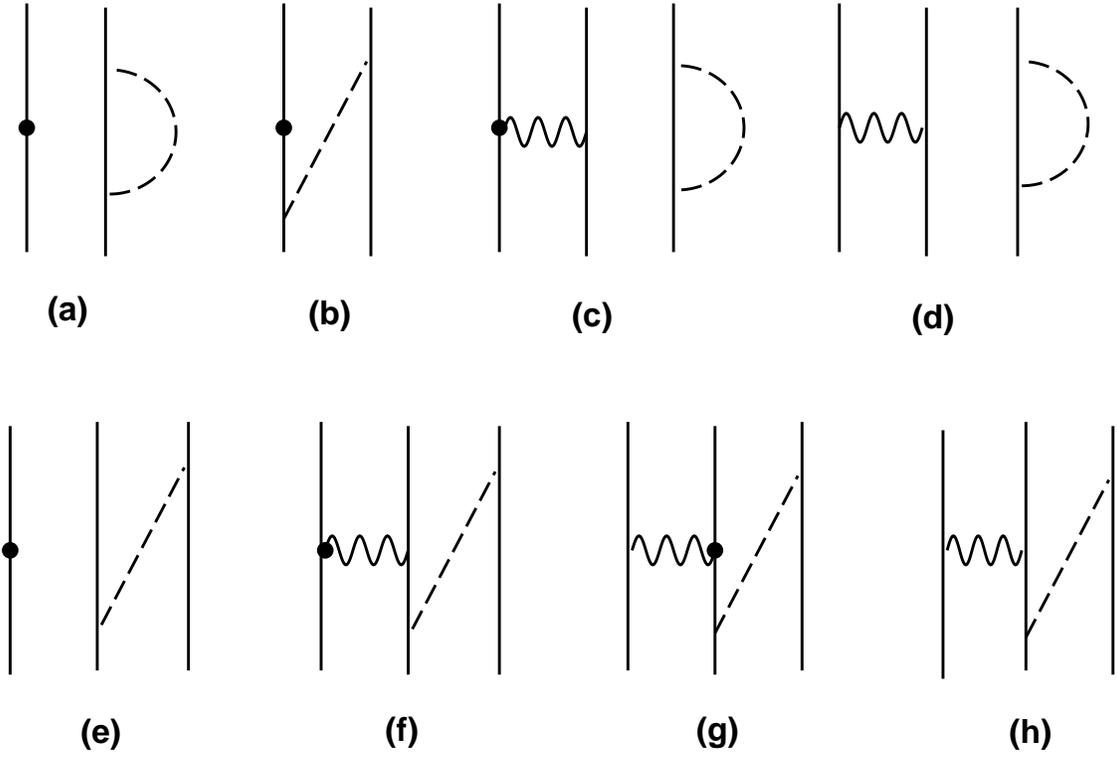}}      
\caption{Time-ordered one-loop graphs which cancel exactly with renormalization graphs.}
\label{fig7}
\end{figure}

The contributions from mass insertions and the renormalization terms cancel exactly for the diagrams in Fig.\ \ref{fig7} as suggested by their topology. In each case, we can move at least one of the loop vertices through the relevant part of factor ${\cal H}_{M,tree}$ to obtain a modified operator expression in which the mass insertion is outside the meson loop and has the topology of a renormalization term. Their sum gives the negative of the corresponding parts of Eq.\ (\ref{mass_renorm}).  

We find only partial cancellations for the diagrams in Figs.\ \ref{fig5} and \ref{fig6}. In these cases, at least one spin or flavor factor in the insertion is trapped inside the loop, and we are left with the structures, labeled by the diagrams,
\begin{eqnarray}
\label{insertion_structures}
{\rm Fig.\ \ref{fig5}\,(a):}\quad && \frac{4}{9}\delta m \sum_lI'_l\,\openone_{j'j} \left(\lambda^l\lambda^l\right)_{k'k} \mbox{\boldmath$\sigma$}_{j} \!\cdot\! \mbox{\boldmath$\sigma$}_{k}, 
 \\
{\rm Fig.\ \ref{fig5}\,(b):}\quad && -\frac{8}{9}\delta m\sum_lI'_l\,\lambda^l_{j'j} \lambda^l_{k'k}\,\mbox{\boldmath$\sigma$}_{j} \!\cdot\! \mbox{\boldmath$\sigma$}_{k}, 
\\
{\rm Fig.\ \ref{fig6}\,(a):}\quad && -\tilde{\alpha}_m \sum_lI'_l\,\left[\left(\lambda^lM\lambda^l\right)_{k'k} - \frac{1}{2}\left(\lambda^l\lambda^lM\right)_{k'k} - \frac{1}{2}\left(M\lambda^l\lambda^l\right)_{k'k}\right], 
\\
{\rm Fig.\ \ref{fig6}\,(c):}\quad && -\tilde{\alpha}_{ss}\sum_lI'_l\,\openone_{j'j}\left[\, \frac{1}{3}\left(\lambda^lM\lambda^l\right)_{k'k} + \frac{1}{2}\left(\lambda^l \lambda^lM\right)_{k'k} + \frac{1}{2}\left(M\lambda^l \lambda^l\right)_{k'k}\right]
\mbox{\boldmath$\sigma$}_{j} \!\cdot\! \mbox{\boldmath$\sigma$}_{k}  
\\
{\rm Fig.\ \ref{fig6}\,(d):}\quad && -\frac{4}{3}\tilde{\alpha}_{ss}M_{j'j}\sum_lI'_l\left(\lambda^l \lambda^l\right)_{k'k}\mbox{\boldmath$\sigma$}_{j} \!\cdot\! \mbox{\boldmath$\sigma$}_{k}, 
\\
{\rm Fig.\ \ref{fig6}\,(e):}\quad &&  \frac{4}{3}\tilde{\alpha}_{ss}\sum_lI'_l\,\left[\left( M\lambda^l + \lambda^lM\right)_{j'j}\lambda^l_{k'k}+\lambda^l_{j'j} \left(M\lambda^l+\lambda^lM\right)_{k'k} \right]\mbox{\boldmath$\sigma$}_{j} \!\cdot\! \mbox{\boldmath$\sigma$}_{k}.
\end{eqnarray}

The explicit forms of the flavor-dependent factors $\sum_lI'_l\,  (\lambda^l\lambda^l)_{k'k}$ and $\sum_lI'_l\,\lambda^l_{j'j}\lambda^l_{k'k}\,\mbox{\boldmath$\sigma$}_{j} \!\cdot\! \mbox{\boldmath$\sigma$}_{k}$ are given in Eqs.\ (\ref{lambda_lambda_sum}) and (\ref{exchange_graphs}), respectively. The remaining flavor factors are 
\begin{eqnarray}
\label{lambda_M_lambda}
\sum_lI'_l\,\left(\lambda^lM\lambda^l\right)_{k'k} &=& 2I'_K\openone_{k'k} - \left(2I'_K-\frac{4}{3}I'_\eta\right)M_{k'k}, 
\\
\label{lambda_lambda_M}
\sum_lI'_l\,\left(\lambda^l\lambda^lM\right)_{k'k} &=& \sum_lI'_l\,\left(M\lambda^l\lambda^l\right)_{k'k} = \left(4I'_K+\frac{4}{3}I'_\eta\right)M_{k'k}, 
\\
\sum_lI'_l\,\left[M_{j'j}\left(\lambda^l\lambda^l\right)_{k'k} + \left(\lambda^l\lambda^l\right)_{j'j}M_{k'k}\right] &=& \left(3I'_\pi+2I'_K+\frac{1}{3}I'_\eta\right) \left(\openone_{j'j}M_{k'k}+M_{j'j}\openone_{k'k}\right)  \nonumber\\
&& -\left(3I'_\pi-2I'_K-I'_\eta\right)M_{j'j}M_{k'k},  
\\
\sum_lI'_l\,\left[\left(M\lambda^l\right)_{j'j}\!\lambda^l_{k'k} + \lambda^l_{j'j}\!\left(M\lambda^l\right)_{k'k}\right] \mbox{\boldmath$\sigma$}_{j} \!\cdot\! \mbox{\boldmath$\sigma$}_{k} &=&  3I'_K\left(\openone_{j'j}M_{k'k}+M_{j'j}\openone_{k'k}\right) \nonumber \\ &=& -\left(I_K'+\frac{2}{3}I_\eta'\right) \left(\openone_{j'j}M_{k'k}+M_{j'j}\openone_{k'k}\right) \,\mbox{\boldmath$\sigma$}_{j} \!\cdot\! \mbox{\boldmath$\sigma$}_{k}  \nonumber \\
&&-I'_K\left(\openone_{j'j}M_{k'k}+M_{j'j}\openone_{k'k}\right)\mbox{\boldmath$\sigma$}_{j} \!\cdot\! \mbox{\boldmath$\sigma$}_{k} \nonumber \\
&&-4 \left(I'_K-I'_\eta\right) M_{j'j}M_{k'k}\,\mbox{\boldmath$\sigma$}_{j} \!\cdot\! \mbox{\boldmath$\sigma$}_{k}.  
\end{eqnarray}

Combining the results for the graphs in Figs.\ \ref{fig5} and \ref{fig6}, including all choices of the quarks involved, we obtain the change in ${\cal H}_{M,tree}$ associated with mass insertions and renormalization at one loop,
\begin{eqnarray}
\label{delta_H_insertions}
\delta{\cal H}_{M,insertions} &=& -6\tilde{\alpha}_mI_\pi'-8\delta mI'_K \nonumber \\
&&+\left[\frac{8}{9}\delta m\left(5I'_\pi+2I'_K\right) - \frac{4}{3}\tilde{\alpha}_{ss}I'_K  \right] \left(\mbox{\boldmath$\sigma$}_{i} \!\cdot\! \mbox{\boldmath$\sigma$}_{j} + \mbox{\boldmath$\sigma$}_{j} \!\cdot\! \mbox{\boldmath$\sigma$}_{k} + \mbox{\boldmath$\sigma$}_{k} \!\cdot\! \mbox{\boldmath$\sigma$}_{i}\right) \nonumber 
\\
&& + \left[\frac{16}{3}\delta m \left(I_\pi'-I_K'\right) + 6\tilde{\alpha}_m I'_K + 16\tilde{\alpha}_{ss}I_K'\right]\left(\,M_i + M_j + M_k\, \right) \nonumber \\
&&-\left[\frac{4}{9}\delta m\left(7I'_\pi-4I'_K-3I'_\eta\right) + \frac{2}{3}\tilde{\alpha}_{ss}\left(6I'_\pi+13I'_K+6I'_\eta \right)\right] \nonumber \\ 
&& \quad \times \left[\,M_i\mbox{\boldmath$\sigma$}_{i} \!\cdot\! \left(\mbox{\boldmath$\sigma$}_{j} + \mbox{\boldmath$\sigma$}_{k}\right) + M_j\mbox{\boldmath$\sigma$}_{j} \!\cdot\! \left(\mbox{\boldmath$\sigma$}_{k} + \mbox{\boldmath$\sigma$}_{i}\right) M_k\mbox{\boldmath$\sigma$}_{k} \!\cdot\! \left(\mbox{\boldmath$\sigma$}_{i} + \mbox{\boldmath$\sigma$}_{j}\right)\, \right]  \nonumber
\\
&&-\left[\frac{8}{9} \delta m \left(I'_\pi-4I'_K+3I'_\eta\right) -8\tilde{\alpha}_{ss}\left(I'_\pi-2I'_K+I'_\eta\right)\right] \nonumber 
\\
&& \quad \times \left[\, M_iM_j \mbox{\boldmath$\sigma$}_{i} \!\cdot\! \mbox{\boldmath$\sigma$}_{j} + M_jM_k \mbox{\boldmath$\sigma$}_{j} \!\cdot\! \mbox{\boldmath$\sigma$}_{k} + M_kM_i \mbox{\boldmath$\sigma$}_{k} \!\cdot\! \mbox{\boldmath$\sigma$}_{i}\,\right],
\end{eqnarray}
where we have suppressed obvious factors of $\openone$. Once again, the terms with zero or one factor of $M$ do not change the structure of the mass operator and can be absorbed by redefinitions of the chiral parameters. The only new terms are those with two factors of $M$. We consider these in the following section.

We consider finally the one-loop tadpole diagram with a quark mass insertion shown in Fig.\ \ref{fig6}\,(b). This diagram arises from the expansion of the expression for ${\cal M}^+$ in Eq.\ (\ref{H_M}) to O($\phi^2$), with the meson fields contracted.  The result reduces to a sum of single-particle operators of the form
\begin{equation}
\label{tadpole_insertion}
\delta{\cal H}_{tad} = -\frac{m}{4f^2}\sum_l\left(\frac{1}{2}\lambda^l \lambda^l M + \lambda^lM\lambda^l +\frac{1}{2}M\lambda^l\lambda^l\right)_{k'k} \int\frac{d^k}{(2\pi)^4}\,\frac{i}{k^2-M_l^2}.
\end{equation}
There is no new chiral structure introduced by these corrections. The integrals that appear involve high loop momenta or short distances. There is no intermediate baryon state to absorb the recoil momenta, and the integrals are only defined after an appropriate but arbitrary ultaviolet regularization. We therefore regard the tadpole terms as short-distance contributions which should be incorporated into the unknown short-distance parameters $m_0$ and $\tilde{\alpha}_m$ in the spirit of chiral perturbation theory. We will assume that this has been done, and will henceforth ignore the tadpole diagram.\footnote{ The redefinition is equivalent to the replacement of ${\cal M}^+$ in Eq.\ (\ref{H_M}) by ${\cal M}^+-\left<0|{\cal M}^+|0\right>$ with the average taken over the meson vacuum.  Given the flavor structure of ${\cal M}^+-\left<0|{\cal M}^+|0\right>$ in Eq.\ (\ref{tadpole_insertion}), this is just a reparametrization of the chiral structure to O($M$).}

\subsection{Expansion in $M$}
\label{subsec:mass-expansion}

We obtain the complete one-loop expression for the baryon mass operator by combining the original expression for ${\cal H}_{M,tree}$ in Eq.\ (\ref{H_M_tree}) with the simple loop corrections $\delta{\cal H}_{M,loop}$ in Eq.\ (\ref{delta_H_M_loop}) and the results for the mass insertions and renormalization terms above. The result is
\begin{eqnarray}
\label{H_M,op}
{\cal H}_{M,op} &=& {\cal H}_{M,tree} + \delta{\cal H}_{M,loop} + \delta{\cal H}_{M,insertions} \nonumber 
\\
&=& \delta m_0 + \frac{1}{3}\delta\hat{m}\,
\left( \,\mbox{\boldmath$\sigma$}_i \!\cdot\! \mbox{\boldmath$\sigma$}_j + \mbox{\boldmath$\sigma$}_j\!\cdot\! \mbox{\boldmath$\sigma$}_k + \mbox{\boldmath$\sigma$}_k\!\cdot\! \mbox{\boldmath$\sigma$}_i\,\right) \nonumber 
\\
&& + \hat{\alpha}_m\left(\,M_i + M_j + M_k\,\right) \nonumber 
\\
&& - \hat{\alpha}_{ss}\left[\,(\mbox{\boldmath$\sigma$}_i+ \mbox{\boldmath$\sigma$}_j)
\!\cdot\! \mbox{\boldmath$\sigma$}_k \,M_k + (\mbox{\boldmath$\sigma$}_k+\mbox{\boldmath$\sigma$}_i)
\!\cdot\! \mbox{\boldmath$\sigma$}_j\,M_j
+ (\mbox{\boldmath$\sigma$}_j+\mbox{\boldmath$\sigma$}_k)
\!\cdot\! \mbox{\boldmath$\sigma$}_i\,M_i\,\right] \nonumber 
\\
&& + \hat{\alpha}_{MM}\,
\left[\, M_iM_j \mbox{\boldmath$\sigma$}_{i} \!\cdot\! \mbox{\boldmath$\sigma$}_{j} + M_jM_k \mbox{\boldmath$\sigma$}_{j} \!\cdot\! \mbox{\boldmath$\sigma$}_{k} + M_kM_i \mbox{\boldmath$\sigma$}_{k} \!\cdot\! \mbox{\boldmath$\sigma$}_{i}\,\right].
\end{eqnarray}
The first four terms in ${\cal H}_{op}$ have the structure of the original tree-level chiral expression in Eq.\ (\ref{H_M}), including an overall shift in the reference mass $m_0$ to
\begin{equation}
\label{delta_m0}
\hat{m}_0 = m_0+\delta m_0 = m_0 -\left(15I_\pi+6I_K+I_\eta\right)-8\delta m I_\pi' - 6\hat{\alpha}_m I'_K.
\end{equation}
This shift can be absorbed by extracting $\hat{m}_0$ instead of $m_0$ in the definition of the heavy-baryon fields, Eqs.\ (\ref{B_v}) and (\ref{T_v}). The other parameters are:
\begin{eqnarray}
\label{delta_m_hat}
\delta\hat{m} &=& \delta m + 2\left(2I_\pi-\frac{1}{3}I_\eta\right)  +\frac{8}{3}\delta m \left(5I'_\pi+2I'_K\right) -4\tilde{\alpha}_{ss}I'_K   
\\
\label{alpha_m_hat}
\hat{\alpha}_m &=& \tilde{\alpha}_m + \left(I_\pi-2I_K+I_\eta\right) + \frac{16}{3}\left(I_\pi'-I_K'\right) +6\tilde{\alpha}_mI'_K + 16\tilde{\alpha}_{ss}I_K', 
\\
\label{alpha_ss_hat}
\hat{\alpha}_{ss} &=& \tilde{\alpha}_{ss} + \frac{2}{3}\left(2I_\pi-I_K-I_\eta\right)  + \frac{4}{9}\left(7I'_\pi-4I'_K-3I'_\eta\right) +\frac{2}{3}\tilde{\alpha}_{ss} \left(6I'_\pi+13I'_K+6I'_\eta\right). 
\end{eqnarray}
The last term in Eq.\ (\ref{H_M,op}), with a coefficient 
\begin{equation}
\label{alpha_MM}
\hat{\alpha}_{MM} = -\frac{2}{3}\left(I_\pi-4I_K+3I_\eta\right) + 8\tilde{\alpha}_{ss} \left(I'_\pi-2I'_K+I'_\eta\right) - \frac{8}{9}\delta m \left(I'_\pi-4I'_K+3I'_\eta\right)
\end{equation}
gives the only new structure. The parameters $\delta m$, $\tilde{\alpha}_m$, and $\tilde{\alpha}_{ss}$ on the right hand sides of these equations are final tree-level parameters determined in fits to the data with the loop corrections included. The parameters that actually appear in the fitted masses are $\hat{m}_0$, $\delta \hat{m}$, $\hat{\alpha}_m$, and $\hat{\alpha}_{ss}$.

We expect the loop contributions to $\hat{m}_0$, $\delta\hat{m}$, $\hat{\alpha}_m$, and $\hat{\alpha}_{ss}$ in Eqs. (\ref{delta_m0})-(\ref{alpha_ss_hat}) to give reasonable estimates of the low-momentum contributions to these quantities when calculated using the expressions in Eqs.\ (\ref{fig4b}) and (\ref{fig4b'}). However, the individual integrals would diverge  in ordinary HBChPT without our imposition of a physical cutoff through the baryon form factor, and the correction terms would then have to be incorporated in redefinitions or renormalizations of the tree-level parameters. 

The mass Hamiltonian in Eq.\ (\ref{H_M,op}) leads to the baryon mass formulas
\begin{eqnarray}
\label{masses}
m_N = \hat{m}_0 - \delta\hat{m}, &\qquad & m_\Delta = \hat{m}_0+\delta\hat{m} ,\nonumber 
\\
m_\Sigma= \hat{m}_0 - \delta\hat{m} + \hat{\alpha}_m + 4\hat{\alpha}_{ss}, &\qquad & m_{\Sigma^*} = \hat{m}_0 + \delta\hat{m} + \left(\hat{\alpha}_m - 2\hat{\alpha}_{ss}\right), \nonumber 
\\
m_\Xi = \hat{m}_0 - \delta\hat{m} + 2\hat{\alpha}_m + 2\hat{\alpha}_{ss} + \hat{\alpha}_{MM}, 
 &\qquad & m_{\Xi^*} = \hat{m}_0 + \delta\hat{m} + 2\left(\hat{\alpha}_m - 2\hat{\alpha}_{ss}\right) + \hat{\alpha}_{MM},  
\\
m_\Lambda = \hat{m}_0 - \delta\hat{m} +\hat{\alpha}_m, &\qquad & m_\Omega =  \hat{m}_0 + \delta\hat{m} + 3\left(\hat{\alpha}_m - 2\hat{\alpha}_{ss}\right) + 3\hat{\alpha}_{MM}. \nonumber
\end{eqnarray}
The octet masses satisfy a modified Gell-Mann--Okubo relation
\begin{equation}
\label{GellMann}
m_\Xi+m_N-\frac{1}{2}\left(3m_\Lambda+m_\Sigma\right) = \hat{\alpha}_{MM},
\end{equation}
where $\tilde{\alpha}_{MM}$ is now a calculated quantity. The equal-spacing rule for the decuplet masses is replaced by the relations
\begin{eqnarray}
\label{equal_spacing}
m_\Omega-2m_{\Xi^*}+m_{\Sigma^*} &=& \hat{\alpha}_{MM}, \nonumber \\
m_{\Xi^*}-2m_{\Sigma^*}+m_\Delta &=& \hat{\alpha}_{MM}, \\
m_\Omega-3m_{\Xi^*}+3m_{\Sigma^*}-m_\Delta &=& 0, \nonumber
\end{eqnarray}
where the third follows from the first two, and states that there are no contributions equivalent to simultaneous insertions of $M$ on all three quark lines.

The relations above, obtained by expanding around the symmetrical limit, predict that $m_\Lambda-m_N-\frac{1}{2}(m_\Sigma-m_\Lambda) = m_{\Sigma*}-m_\Delta$. The two sides of the equation have the experimental values 138 MeV and 153 MeV, respectively, so the relation holds only approximately. The measured values of $\hat{\alpha}_{MM}$ of $-13.1\pm 0.4$ MeV from Eq.\ (\ref{GellMann}), and $-3.8\pm 2.1$ MeV from the second of Eqs.\ (\ref{equal_spacing}), are also different. As we noted at the end of Sec.\ \ref{subsec:mass_operators}, we could use projection operator methods to allow for differences between the values of the matrix elements $\tilde{\alpha}_M$ ($\tilde{\alpha}'_M$) and $\tilde{\alpha}_{ss}$ ($\tilde{\alpha}'_{ss}$) in the octet (decuplet) states. 
The expressions given above for the loop contributions to the various parameters would then change in detail, though the basic structure is fixed. We have not carried through a detailed calculation as the results are already subsumed, though in a less than transparent way, in the general expressions given in  \cite{DH-masses}, \cite{phuoc-diss} and \cite{jaczko-diss}.

\subsection{Momentum structure of the loop corrections}
\label{subsec:momentum_structure}

Chiral perturbation theory is based on a low-momentum expansion, with the high-momentum components of the underlying field theory integrated out and replaced by short-distance parameters in the effective field theory, here $m_0$, $\delta m$, $\tilde{\alpha}_m$, and $\tilde{\alpha}_{ss}$. The meson loop corrections are then supposed to introduce the calculable effects of interactions at low momenta or long distances. However, calculations based on the pseudoscalar mesons alone will not be reliable unless the dominant momenta in the loops are low on the scale of the vector meson masses or the chiral cutoff, generally taken as about 1 GeV. For momenta on these scales,  the vector mesons come into play and the pseudoscalar octet should be extended to the full {\bf 35} of SU(6). Multiloop and other short-distance effects may also become important. We therefore think it is important to examine the momentum structure of the loop integrals.

We have used the dipole form factor
\begin{equation}
\label{F}
F(k^2)=1/\left(1+k^2/\lambda^2\right)^2
\end{equation}
and the general expressions in \cite{DH-masses} in our calculations. We obtain a very good fit to the baryon masses, with an average deviation from experiment of 0.6 MeV, for $\lambda=930$ and the SU(6) couplings $\beta=D=0.75$, $D+F=1.25\approx |g_A/g_V|$, and $f_\pi=93$ MeV. The form factor is close to the measured electromagnetic form factor of the proton which has the same form with $\lambda=850$ MeV, so we regard the fitted value of $\lambda$ as reasonable.  

\begin{figure}
\def\picsize{6in}                  
\def\picfilename{fig8.eps}         
\centerline{\epsfxsize\picsize\epsfbox{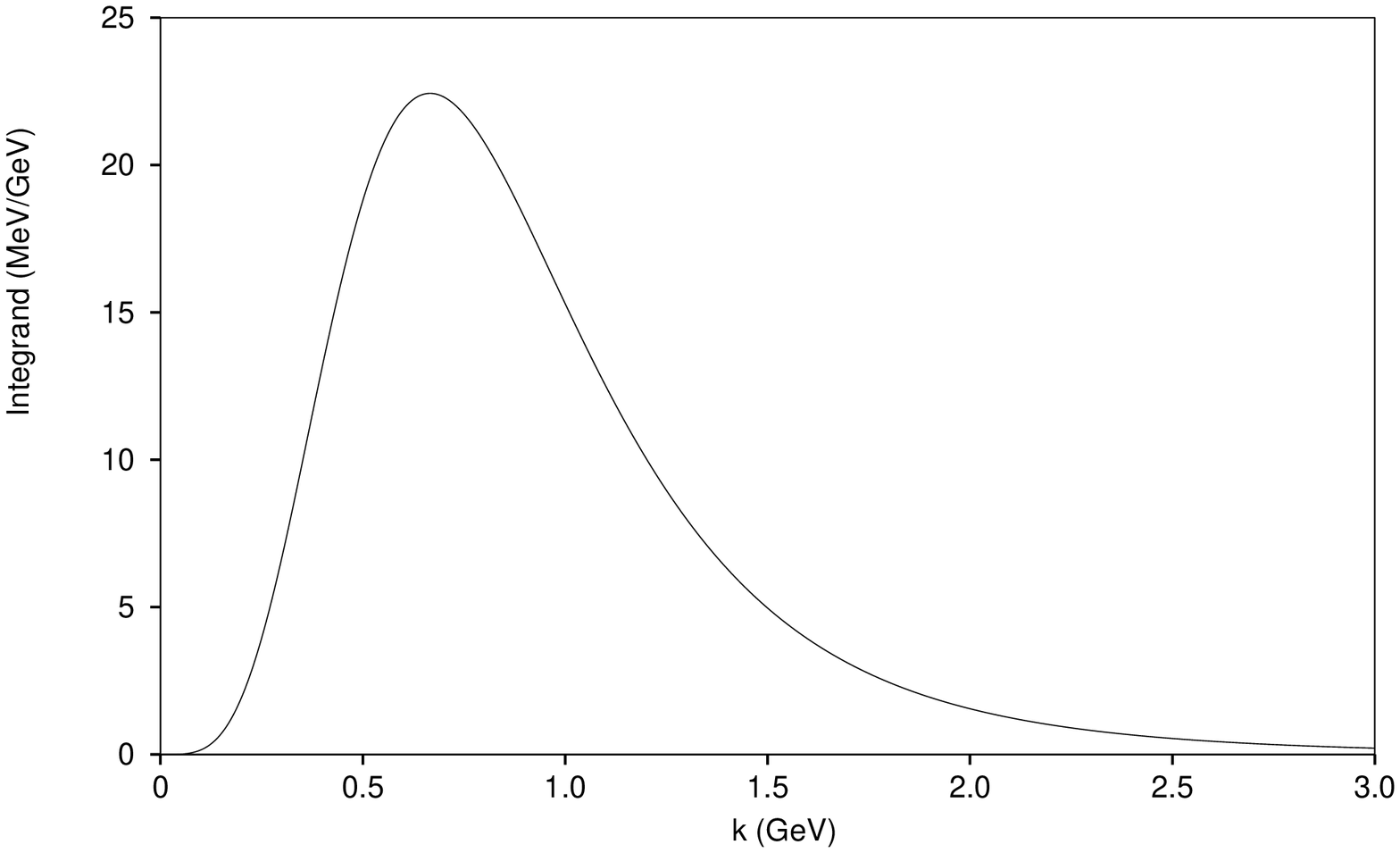}}      
\caption{Momentum dependence of the combination of integrals $-\frac{2}{3}(I_\pi-4I_K+3I_\eta)$ which appears in  $\hat{\alpha}_{MM}$ in the 
baryon mass formula in Eq.\ (\protect\ref{masses}).}
\label{fig8}
\end{figure}

In Fig.\ \ref{fig8}, we show the momentum structure of the resulting integrand for  $I_K$. The peak in the integrand is at a momentum of 700 MeV, somewhat high on the scale of $M_K$, and there are clearly significant contributions to the integral from momenta  above the supposed chiral cutoff around 1 GeV. These high-momentum contributions to $I_K$ are certainly not reliable. The same is true for the integrals $I_\pi$ and $I_\eta$. The integrands for $I'_K$, $I_\pi$, and $I'_\eta$ are shifted toward somewhat lower values of $k$, but the integrals still contain significant contributions from momenta above the chiral cutoff. While the resulting high-momentum contributions in Eqs.\ (\ref{delta_m0})-(\ref{alpha_ss_hat}) can be absorbed formally by redefinitions of $\delta m$ and $\tilde{\alpha}$,\footnote{The quantities on the left hand sides of Eqs.\ (\ref{delta_m_hat})-(\ref{alpha_ss_hat}) are fixed by the data. We therefore have a set of linear equations for $\delta m$, $\tilde{\alpha}_m$ and $\tilde{\alpha}_{ss}$ which can be solved whatever the treatment of the high-momentum components of the integrals.} this requires the introduction of some new, undefined, cutoff procedure. We therefore regard our calculated loop contributions in those equations only as plausible estimates, with the ``low momentum'' contributions overestimated. The difficulties with the high-momentum contributions have also been discussed by Donoghue, Holstein, and Borasoy \cite{Don-masses} from a somewhat different point of view.

\begin{figure}
\def\picsize{6in}                  
\def\picfilename{fig9.eps}         
\centerline{\epsfxsize\picsize\epsfbox{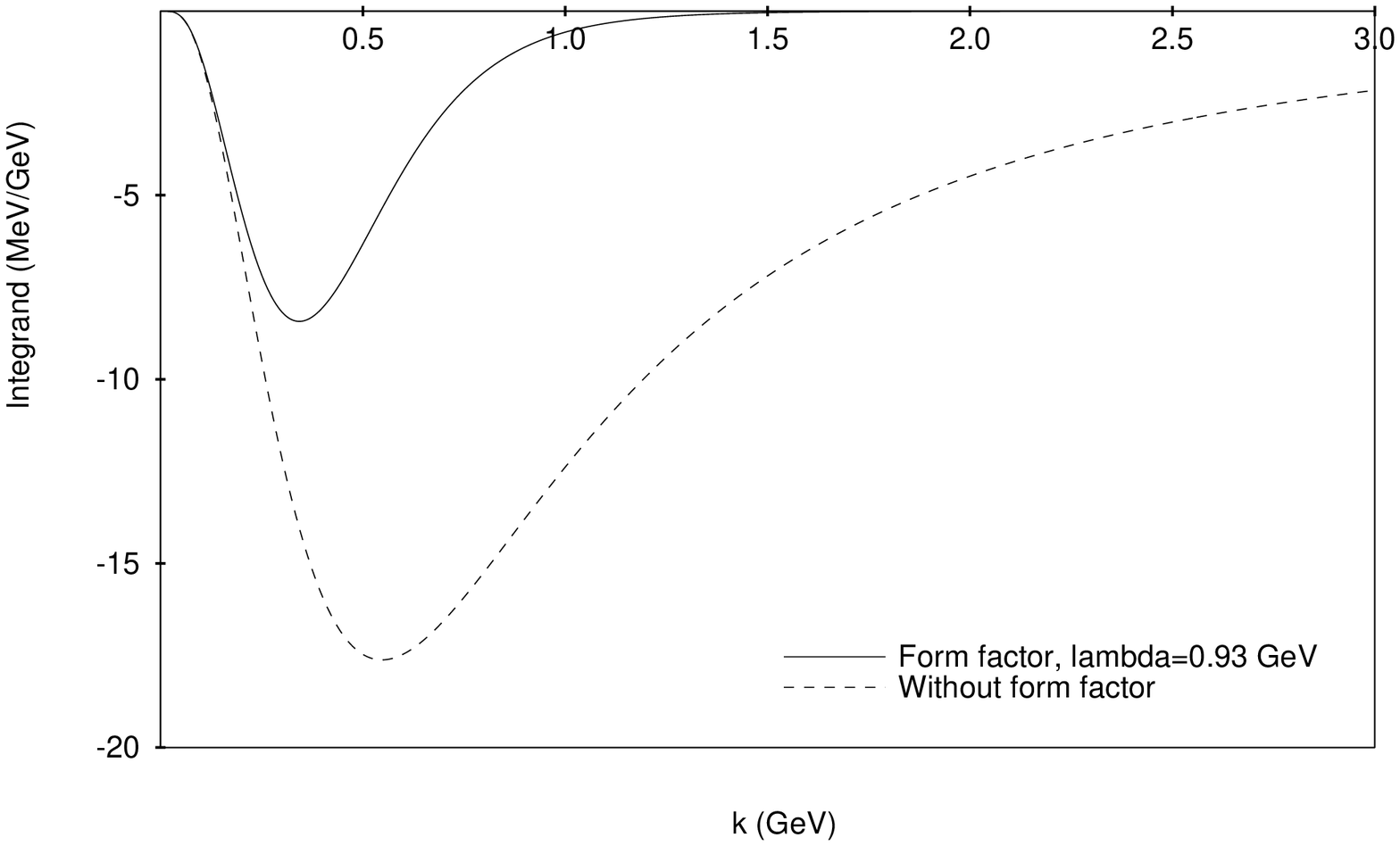}}      
\caption{Momentum dependence of the combination of integrals $-\frac{2}{3}(I_\pi-4I_K+3I_\eta)$ which appears in  $\hat{\alpha}_{MM}$ in the 
baryon mass formula in Eq.\ (\protect\ref{masses}).}
\label{fig9}
\end{figure}

The situation is much better for the new parameter $\hat{\alpha}_{MM}$ defined in Eq.\ (\ref{alpha_MM}). The combinations of integrals in the first and third terms on the right hand side of that equation are convergent for meson masses which satisfy the Gell-Mann--Okubo mass formula $M_\pi^2-4M_K^2+3M_\eta^2=0$.
In Fig.\ \ref{fig9}, we show the momentum structure of the term $-\frac{2}{3}\left(I_\pi-4I_K+3I_\eta\right)$ calculated using the form factor above. As is evident from the figure, most of the contribution to $\hat{\alpha}_{MM}$ comes from momenta in the range of the $K$ and $\eta$ masses, significantly below the dominant momentum ranges in the individual integrals. This is ``low'' on the chiral scale, and we therefore expect the result to be reliable. 

We show for comparison the integrand for $-\frac{2}{3}\left(I_\pi-4I_K+3I_\eta\right)$ which we obtain by omitting the form factor. This integrand contains substantial high-momentum components. The high-momentum components of the resulting integral cannot be absorbed in this case in a redefinition of a short-distance parameter, and the result for $\hat{\alpha}_{MM}$, a supposed long-range or low-momentum parameter, is overestimated. The presence of the form factor is both important for consistency with a low-momentum expansion, and motivated physically by the extended structure of the baryons. It corresponds in the chiral context to a selective summation of higher order terms in the derivative expansion. 

The integrand for the last term on the right hand side of Eq.\ (\ref{alpha_MM}) has the same behavior, is concentrated at low momenta when the form factor is included, and leads to a convergent integral even when it is omitted provided the meson masses satisfy the Gell-Mann--Okubo formula. The second term has different coefficients for the $I'$\,s, and gives a logarithmically divergent integral without the form factor. However, the dominant contributions again come from low momenta when the form factor is included, and we conclude that the result should be reliable.

A different question concerns the adequacy of the parametrization of the baryon masses given by Eqs.\ (\ref{masses}) with the octet and decuplet parameters allowed to differ. This is measured by the modified Gell-Mann--Okubo relations in Eqs.\ (\ref{GellMann}) and (\ref{equal_spacing}). The calculated value of $\hat{\alpha}_{MM}$ for the baryon octet is -14.0 MeV for the fit described above. This is to be compared to the experimental value $\hat{\alpha}_{MM} = -13.1\pm 0.4$ MeV determined by the left hand side of Eq.\ (\ref{masses}). The agreement is good.  

The first two decuplet mass relations in Eq.\ (\ref{equal_spacing}) give experimental values of $\hat{\alpha}'_{MM}$ of $-9.7\pm 0.07$ and $-3.8\pm 2.1$  MeV, respectively. The fact that the values are different and that the third relation is not satisfied indicates the presence of a term cubic in $M$ in the mass relation, with a coefficient $\hat{\alpha}'_{MMM} = m_\Omega-3m_{\Xi^*} + 3m_{\Sigma^*}- m_\Delta = -5.9\pm 2.2$ MeV. This is similar in magnitude to the value $\hat{\alpha}'_{MM}=-3.8\pm 2.1$ MeV determined by the second relation, which is independent of the invariant $MMM$, and to the value $\hat{\alpha}'_{MM} = -7.3$ MeV obtained in our fit. This clearly indicates the presence of significant contributions to the masses beyond those encompassed in the meson loop corrections considered here. The fitted value of $\hat{\alpha}'_{MM}$ roughly averages the experimental values from the first two decuplet relations presumably to reduce the residual error in the fit in the absence of a cubic term $\hat{\alpha}'_{MMM}$ in the theory. We are not aware of a low-order source for an $\hat{\alpha}'_{ss}$ term that does not involve new parameters.

\section{LOOP CORRECTIONS TO THE BARYON MAGNETIC MOMENTS}
\label{sec:moments}

\begin{figure}
\centerline{
\epsffile{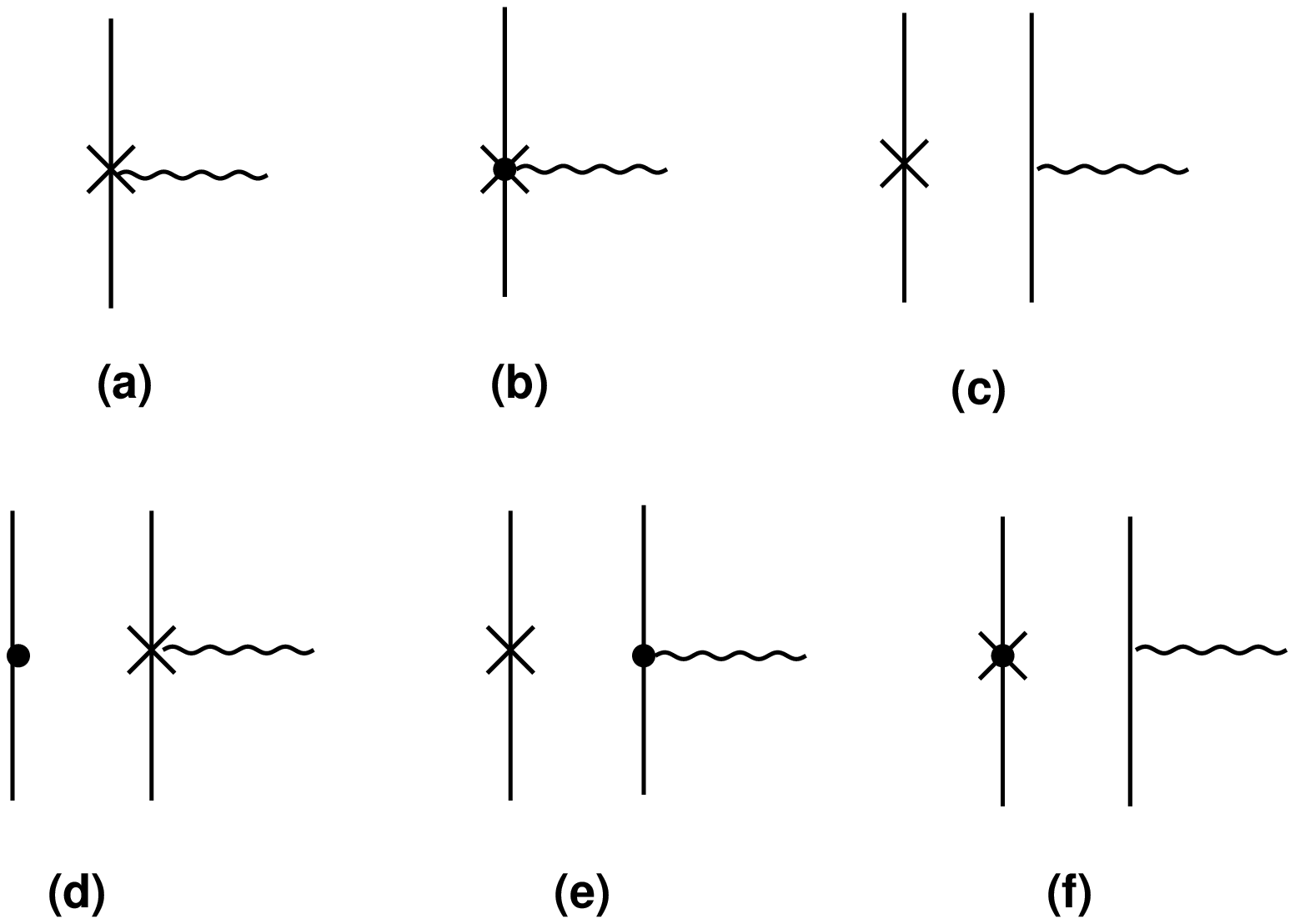}
}
\caption{One- and two-body contributions to the baryon magnetic moments at the quark level. Lines with small wiggles represent photons. There is a factor $\mbox{\boldmath$\sigma$}\!\cdot\!{\bf B} \leftrightarrow -\sigma^{\mu\nu} F_{\mu\nu}$ in the Lagrangian at the quark-photon vertex, where $\bf B$ is the external magnetic field. Crosses correspond to factors of the quark charge matrix $Q$, and dots to insertions of the flavor matrix $M$. Graphs (a) and (b) give the SU(6) structure for the moments.}
\label{fig10}
\end{figure}

\begin{figure}
\centerline{
\epsffile{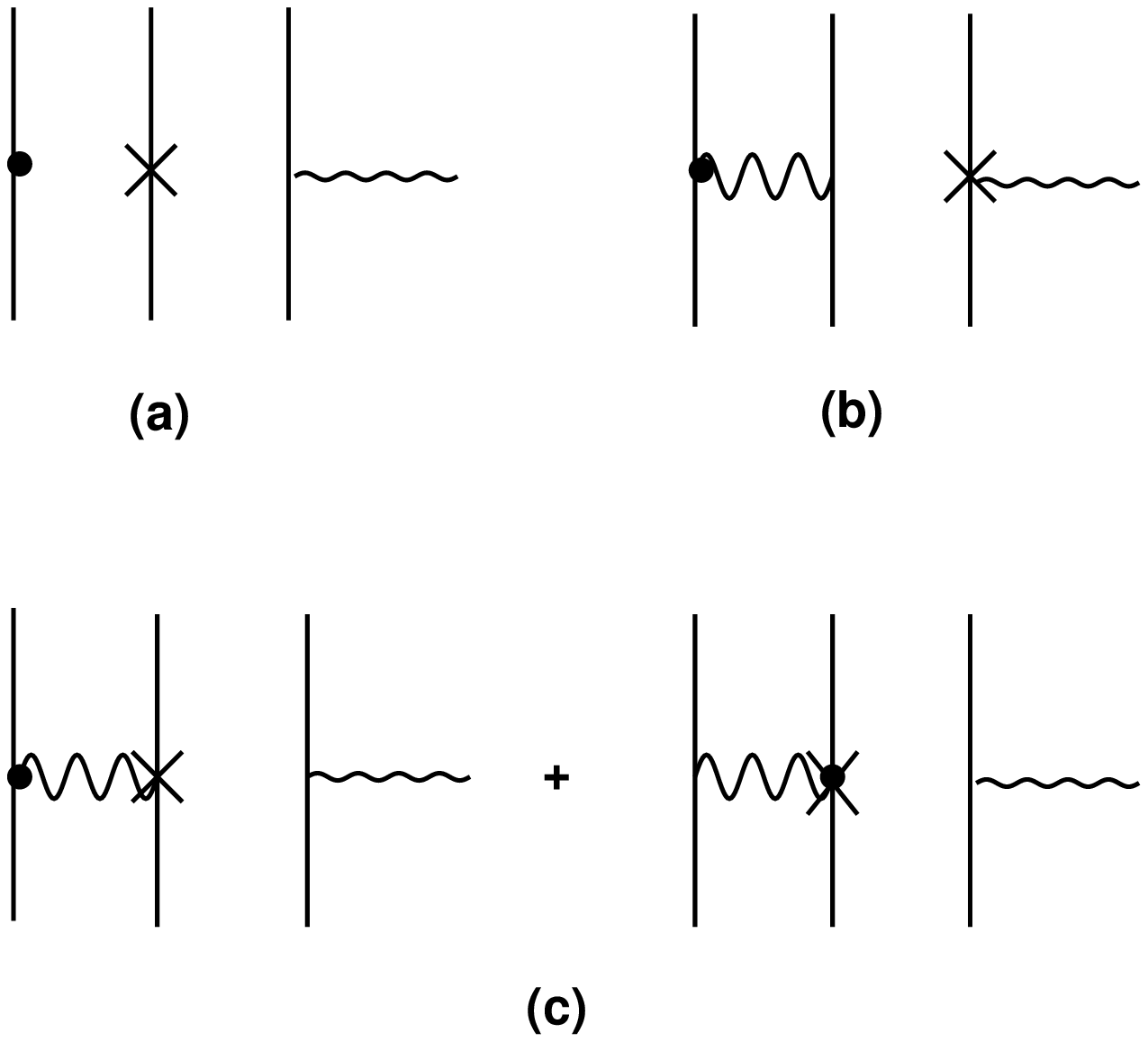}
}
\caption{Three-body contributions to the baryon magnetic moments. The two graphs in (i) appear together with the same overall coefficient. The structures in graphs (h) and (i) are reducible to those in Figs.\ \protect\ref{fig10} and \protect\ref{fig11}\,(g).}
\label{fig11}
\end{figure}

In \cite{DHJ}, we analyzed the structure of the baryon magnetic moment operators to O($m_s$), connecting the general spin-flavor structure to the spin dependence of the underlying dynamical theory. The structure is rather complex, with the Lagrangian for the octet moments involving both the one- and two-body quark-level operators
\begin{eqnarray}
\label{one-body_moments}
&& (Q\,\mbox{\boldmath$\sigma$})_k\!\cdot\!{\bf B},\qquad (QM\,\mbox{\boldmath$\sigma$})_k\!\cdot\!{\bf B}, \\ 
\label{two-body_moments}
Q_j\,\mbox{\boldmath$\sigma$}_k\!\cdot\!{\bf B}, \qquad
&& M_j(Q\,\mbox{\boldmath$\sigma$})_k\!\cdot\!{\bf B}, \qquad Q_j(M\,\mbox{\boldmath$\sigma$})_k\!\cdot\!{\bf B}, \qquad (QM)_j\,\mbox{\boldmath$\sigma$}_k\!\cdot\!{\bf B},
\end{eqnarray}
shown diagrammatically in Fig.\ \ref{fig10}, and the three-body operators
\begin{eqnarray}
\label{three-body_moments}
&& M_iQ_j\,\mbox{\boldmath$\sigma$}_k\!\cdot\!{\bf B}, \qquad
M_iQ_k\,\mbox{\boldmath$\sigma$}_i\!\cdot\!\mbox{\boldmath$\sigma$}_j \,\mbox{\boldmath$\sigma$}_k\!\cdot\!{\bf B}, \qquad (M_iQ_j+Q_iM_j)\, \mbox{\boldmath$\sigma$}_i\!\cdot\!\mbox{\boldmath$\sigma$}_j \,\mbox{\boldmath$\sigma$}_k\!\cdot\!{\bf B}
\end{eqnarray}
shown in Fig.\ \ref{fig11}. $\bf B$ is the magnetic field, the subscripts indicate the quarks on which the respective operators act, and the labels are to be assigned in all possible ways. The same structures appear for the decuplet moments and the octet-decuplet transition moments, but with potentially different coefficients \cite{DHJ}, and with $\mbox{\boldmath$\sigma$}_i \!\cdot\! \mbox{\boldmath$\sigma$}_j \rightarrow 1$ for decuplet states.

The one-body terms in Fig.\ \ref{fig10}\,(a) and (b) combine to give the additive quark model for the moments with an effective moment operator $\mu_1Q+\mu_2QM$, and are the dominant terms.\footnote{As noted in \cite{DHJ}, the one-body moment $\mu_i$ in dynamical models corresponds to the matrix element of $q_i/2E_i=q_i/2\sqrt{{\bf p}_i^2+m_i^2}$. Because $M^n=M$, the effective moment operator $\mu_{i'i}=(\mu_1 Q+\mu_2 QM)_{i'i}$ above actually sums all orders in the expansion of $1/E_i$ in powers of $m_s$. The quark model moments, though represented as $\mu_1 Q+\mu_2 QM$, are therefore not restricted to first-order corrections in $m_s$. The additional effects of $m_s$ on the wave functions lead in the matrix elements $\left<q_i/2E_i\right>$ to the extra corrections described in first order by the diagram in Fig.\ \ref{fig10}\,(d).} Graph (c) gives a Thomas-type contribution to the moments as discussed in \cite{DHJ} and \cite{DH-loop-moments}.  Graph (d) describes the two-body contribution of the strange-quark mass to the leading term (a) at O($m_s$), while (e) and (f) give the corresponding effects on the Thomas term. Graph (g) in Fig.\ \ref{fig11} gives the three-body correction to the Thomas term, while (h) and (i) arise from the effect of spin-spin interactions on the leading and Thomas terms at O($m_s$).

When restricted to the baryon octet, the nine operators above can be related to the nine independent chiral invariants
\begin{eqnarray}
\label{chiral_terms}
&&{\rm Tr}\,\bar{B}QB, \quad {\rm Tr}\,\bar{B}BQ, \quad {\rm Tr}\,\bar{B}QMB, \quad {\rm Tr}\,\bar{B}BQM, \quad {\rm Tr}\,\bar{B}QBM, \nonumber \\ && {\rm Tr}\,\bar{B}MBQ, \quad ({\rm Tr}\,M)\,({\rm Tr}\,\bar{B}QB), \quad \quad ({\rm Tr}\,M)\,({\rm Tr}\,\bar{B}BQ), \quad ({\rm Tr}\,MQ)\,({\rm Tr}\,\bar{B}B),
\end{eqnarray}
in standard treatments of HBChPT \cite{Jen-moments,DHJ,Bos}. For example, the combination of the graphs (a) and (c) with no mass insertions can be reduced to the Coleman-Glashow  SU(3)-symmetric form for the moments, with the Lagrangian \cite{Coleman}
\begin{equation}
\label{coleman-glashow}
{\cal L}_\mu = {e \over {4 m_N}} \left( \mu_D {\rm Tr} \, \bar B \{ Q, B \} + \mu_F {\rm Tr} \, \bar B [ Q, B ] \right).  
\end{equation}
We have suppressed a factor $\sigma^{\mu\nu}F_{\mu\nu}$ which acts on the field $B$ in writing this and the preceding expressions. 

Two of the invariants in Eq..\ (\ref{chiral_terms}) are simply products of ${\rm Tr}\,M$ with other invariants and cannot be separated from them, so there are seven independent parameters to be used in fitting the octet moments. Using the methods of \cite{DHJ}, it is, in fact, straightforward to show that the structures in Fig.\ \ref{fig11}\,(h) and (i) can be expressed in terms of the other structures in Figs.\ \ref{fig10} and \ref{fig11}. We will take advantage of this reduction, rearrange the remaining terms, and write the moment operator as a sum of the operators
%
%\begin{mathletters}
\begin{eqnarray}
\label{basic}
{\bf m}_1 &=& \left(Q_i\mbox{\boldmath$\sigma$}_i + Q_j\mbox{\boldmath$\sigma$}_j  +  Q_k\mbox{\boldmath$\sigma$}_k\right) = \sum_l\,Q_l\mbox{\boldmath$\sigma$}_l,  \\
\label{QM}
{\bf m}_2 &=& \left(Q_iM_i\mbox{\boldmath$\sigma$}_i + Q_jM_j\mbox{\boldmath$\sigma$}_j  +  Q_kM_k\mbox{\boldmath$\sigma$}_k\right) = \sum_l \,(QM)_l\mbox{\boldmath$\sigma$}_l =Q_s \sum_l\,M_l\mbox{\boldmath$\sigma$}_l,  \\
\label{Qsigma}
{\bf m}_3 &=& \left(Q_i + Q_j + Q_k\right)\left(\mbox{\boldmath$\sigma$}_i + \mbox{\boldmath$\sigma$}_j  +  \mbox{\boldmath$\sigma$}_k\right) = Q\mbox{\boldmath$\sigma$}, \\
\label{M*basic}
{\bf m}_4 &=& \left(M_i + M_j + M_k\right)\,\sum_l \,Q_l\mbox{\boldmath$\sigma$}_l  = N_s\,\sum_l \,Q_l\mbox{\boldmath$\sigma$}_l ,  \\
\label{Q*Msigma}
{\bf m}_5 &=& \left(Q_i + Q_j + Q_k\right)\left(M_i\mbox{\boldmath$\sigma$}_i + M_j\mbox{\boldmath$\sigma$}_j + M_k\mbox{\boldmath$\sigma$}_k\right) = Q\,\sum_lM_l\mbox{\boldmath$\sigma$}_l , \\
\label{Q_sN_s}
{\bf m}_6 &=& \left(Q_iM_i + Q_jM_j + Q_kM_k\right)\left(\mbox{\boldmath$\sigma$}_i + \mbox{\boldmath$\sigma$}_j + \mbox{\boldmath$\sigma$}_k\right) = Q_sN_s\mbox{\boldmath$\sigma$}, \\
\label{QN_s}
{\bf m}_7 &&  \left(Q_i+Q_j+Q_k\right)\left(M_i+M_j+M_k\right) \left(\mbox{\boldmath$\sigma$}_i+\mbox{\boldmath$\sigma$}_j +\mbox{\boldmath$\sigma$}_k\right) = QN_s\mbox{\boldmath$\sigma$},
\end{eqnarray}
%\end{mathletters}
%
with $\mbox{\boldmath$\mu$}=\sum_l\mu_l{\bf m}_l$. The labels $i,\,j,\,k$ on $Q$, $M$, and $\mbox{\boldmath$\sigma$}$ in these expressions denote the quark on which the operator acts so that, for example, $Q_iM_j\mbox{\boldmath$\sigma$}_j$ corresponds to the matrix element $\bar{B}_{k'j'i'}Q_{i'i} M_{j'j} \mbox{\boldmath$\sigma$}_j \openone_{k'k}B_{ijk}$. The sum $Q=Q_i+Q_j+Q_k$ is the total charge operator, $N_s=M_i+M_j+M_k$ measures the number of strange quarks in the baryon, and $\mbox{\boldmath$\sigma$} = \mbox{\boldmath$\sigma$}_i + \mbox{\boldmath$\sigma$}_j  +  \mbox{\boldmath$\sigma$}_k$ is the total spin operator and acts only on the free spinor index of $B_{ijk}^{\,\gamma}$ or $T_{ijk}^{\,\mu,\gamma}$. The coefficients for the octet moments are given in Table \ref{table:octet_moments}. The moments automatically satisfy the Okubo sum rule
\begin{equation}
\label{Okubo_sum}
\mu_{\Sigma^+}+\mu_{\Sigma^-}-4\left(\mu_n+\mu_{\Xi^0}\right)+6\mu_\Lambda - 4\sqrt{3}\mu_{\Sigma^0\Lambda} = 0,
\end{equation}
valid to first order in $M$.

If we rewrite $\mbox{\boldmath$\mu$}$ in terms of operators ${\bf O}_r$ which correspond to the the diagrams in Fig.\ \ref{fig10}\,(a)-(f) and Fig.\ \ref{fig11}\,(g) including all choices of the quarks, we get the alternative expression $\mbox{\boldmath$\mu$}=\sum_{r=a}^g \mu_r {\bf O}_r$, where
\begin{eqnarray}
\label{moments_a}
\mu_a &=& \mu_1+\mu_3, \nonumber \\
\mu_b &=& \mu_2+\mu_4+\mu_5, \nonumber \\
\mu_c &=& \mu_3 ,\nonumber \\
\mu_d &=& \mu_4+\mu_6,  \\
\mu_e &=& \mu_5+\mu_7, \nonumber \\
\mu_f &=& \mu_6+\mu_7, \nonumber \\
\mu_g &=& \mu_7. \nonumber 
\end{eqnarray}
This has the advantage that the various terms have simple dynamical interpretations \cite{DHJ}, thus providing more insight into their likely magnitudes. However, the expressions for the baryon moments become somewhat more complicated.

An analysis of the moments in terms of any of these sets of parameters gives a very good fit to the data as discussed, for example, in \cite{Bos}. However, as we emphasized elsewhere \cite{DH-ChPTmoments}, the fit is determined largely by the seven precisely measured moments for the $p$, $n$, $\Sigma^+$,  $\Sigma^-$,$\Lambda$, $\Xi^0$ and $\Xi^-$, which could be matched exactly in the absence of the poorly measured $\Sigma^0$-$\Lambda$ transition moment. In this sense, the fit is probably best regarded as giving a prediction of the $\Sigma^0$-$\Lambda$ transition moment, with $\Sigma^0$ moment determined by isospin invariance.\footnote{A fit to the moments other than $\mu_{\Sigma^0\Lambda}$ using the parameters $\mu_1$-$\mu_7$ gives the values $\mu_1= 2.870$, $\mu_2=-1.013$, $\mu_3=-0.077$, $\mu_4=-0.3015$, $\mu_5=0.086$, $\mu_6=0.284$, and $\mu_7=0.145$, all in units of the nuclear magneton. Alternatively, using the parameter $\mu_a$-$\mu_g$, we obtain the values the values $\mu_a= 2.793$, $\mu_b=-1.400$, $\mu_c=-0.077$, $\mu_d=-0.018$, $\mu_e=0.060$, $\mu_f=0.429$, and $\mu_g=0.145$. The Okubo sum rule predicts that $\mu_{\Sigma^0\Lambda}=1.48$ in either case, while the experimental value is $1.61\pm 0.08$. We note that the largest parameters in the second fit are $\mu_a$, $\mu_b$, and $\mu_f$. The first two give the additive quark model. The third gives the nonadditive Thomas-type contribution involving the strange quark. Dynamical estimates \cite{phuoc-diss,DHJ} predict small values of $\mu_c$, $\mu_d$, and $\mu_e$ in agreement with the fit.}

A number of authors have calculated loop corrections to the moments \cite{Caldi,Gasser,Krause,Jen-moments,Luty,Daietal,Mei-moments,%
DH-ChPTmoments,DH-loop-moments,Ha-moments}. It is clear from their results that the bulk of the corrections can be absorbed in adjustments of the original short-distance chiral parameters \cite{phuoc-diss} as we showed above for the baryon masses.  However, because of the large number of chiral parameters and the limited experimental information available,\footnote{In \cite{DH-loop-moments}, two of us took a dynamical approach, calculated the Thomas-type corrections approximately, included the loop corrections, and  fitted the moment data using only the quark moments $\mu_u=-2\mu_d$, $\mu_s$, and the cutoff $\lambda$ in the form factor as adjustable parameters. The resulting three-parameter fit was excellent, with a mean deviation of theory from from experiment of 0.05 $\mu_N$ compared to 0.12 $\mu_N$ for the additive quark model. It also predicted the the values of the $\Omega^-$ moment and the $\Sigma^0-\Lambda$ transition moment within their experimental uncertainties. We do not believe much, if any, insight would be gained by including all the chiral parameters in the fit. What is needed is improvement in the dynamical calculations, including estimates of the extra parameters.} we will simply sketch the argument for the baryon octet, and will concentrate on the structure of most important corrections. As in the case of the baryon masses, we would like to extract the new structures introduced by the meson loop corrections. 

\begin{figure}
\centerline{
\epsffile{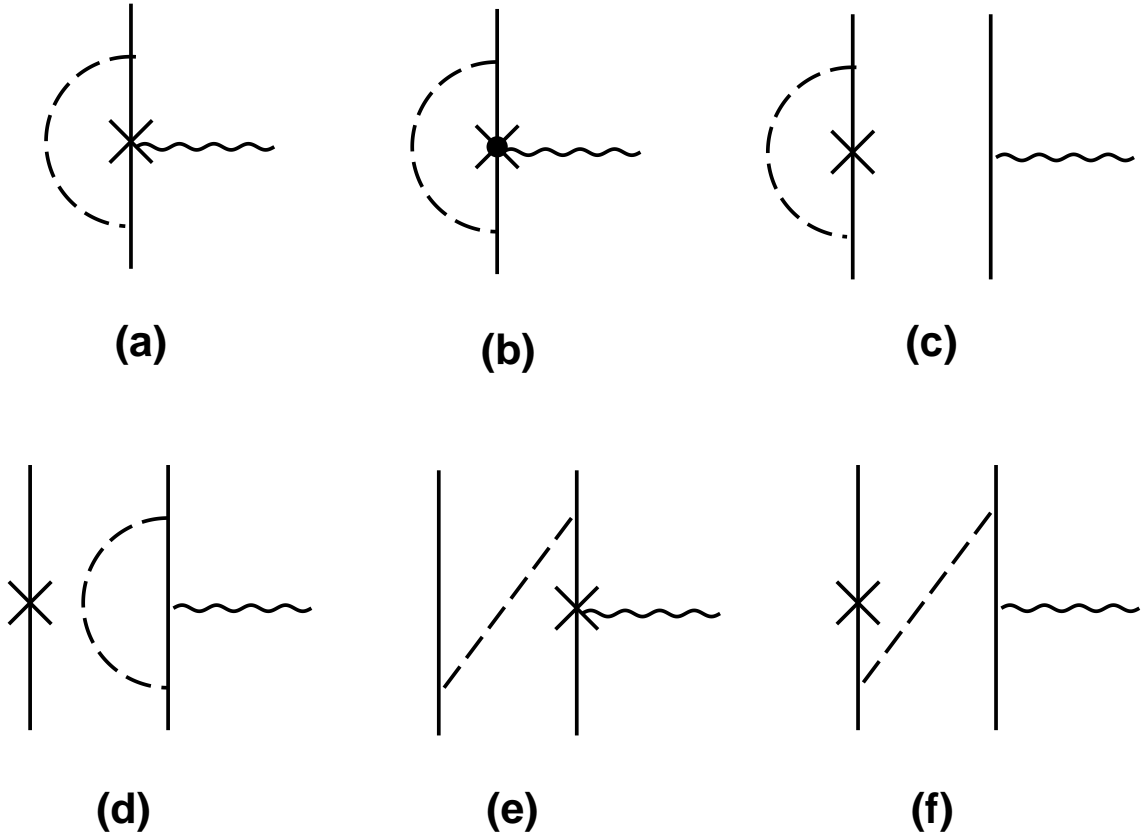}
}
\caption{Typical meson loop corrections to the leading contributions to the baryon magnetic moments. (a) and (b): corrections to the additive moments. (c) and (d): corrections to the Thomas terms. (e): a graph that cancels identically with the renormalization terms. (f): a nontrivial exchange graph. }
\label{fig12}
\end{figure}

\begin{figure}
\centerline{
\epsffile{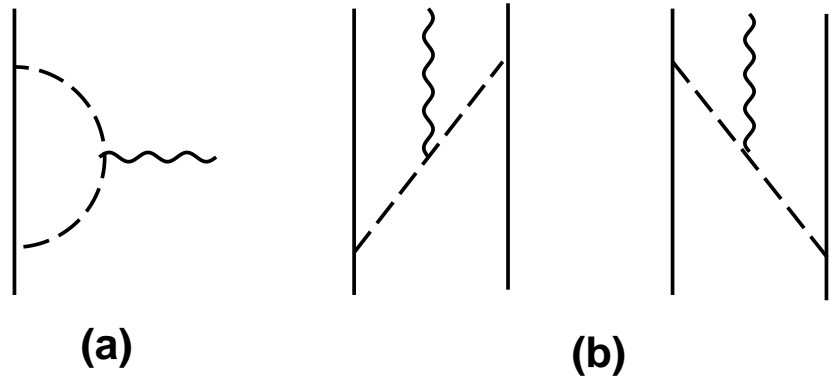}
}
\caption{Time-ordered graphs which involve direct meson-photon interactions. }
\label{fig13}
\end{figure}

We show typical loop diagrams in Figs.\ \ref{fig12} and  \ref{fig13}. These show, respectively, the corrections to the leading quark-model and Thomas-type contributions to the moments, and the new contributions that arise directly from the meson currents.  Fig.\ \ref{fig12}\,(e) involves no ``trapped'' vertex, and cancels exactly with the corresponding wave function renormalization diagram. The remaining contributions involve the structures, labelled by the diagrams, 
\begin{eqnarray}
\label{10a}
{\rm Fig.\ \ref{fig12}\,(a):}\quad && -\frac{1}{3}\sum_lI'_l\,\left(\lambda^l Q\lambda^l\right)_{k'k} \mbox{\boldmath$\sigma$}_k\!\cdot\!{\bf B},
\\
\label{10b}
{\rm Fig.\ \ref{fig12}\,(b):}\quad && -\frac{1}{3}\sum_lI'_l\,\left(\lambda^l MQ\lambda^l\right)_{k'k} \mbox{\boldmath$\sigma$}_k\!\cdot\!{\bf B}, 
\\
\label{10c}
{\rm Fig.\ \ref{fig12}\,(c):}\quad && \sum_lI'_l\,\left(\lambda^l Q\lambda^l\right)_{j'j}\openone_{k'k}\; \mbox{\boldmath$\sigma$}_k \!\cdot\! {\bf B}, 
\\
\label{10d}
{\rm Fig.\ \ref{fig12}\,(d):}\quad && -\frac{1}{3}Q_{j'j}\sum_lI'_l\,\left(\lambda^l\lambda^l\right)_{k'k} \mbox{\boldmath$\sigma$}_k\!\cdot\!{\bf B}, 
\\
\label{10f}
{\rm Fig.\ \ref{fig12}\,(f):}\quad && -\frac{1}{3}\sum_lI'_l\, \left(Q\lambda^l\right)_{j'j}\lambda^l_{k'k}\, \left(\mbox{\boldmath$\sigma$}_j \!\cdot\! \mbox{\boldmath$\sigma$}_k\right) \left(\mbox{\boldmath$\sigma$}_k\!\cdot\!{\bf B}\right), 
\\
\label{11a}
{\rm Fig.\ \ref{fig13}\,(a):}\quad && \frac{1}{3}\sum_l Q_lJ_l\,\left(\lambda^l \lambda^l\right)_{k'k} \mbox{\boldmath$\sigma$}_k\!\cdot\!{\bf B}, 
\\
\label{11b}
{\rm Fig.\ \ref{fig13}\,(b):}\quad &&  \frac{1}{3} \sum_l Q_lJ_l\; \lambda^l_{j'j}\lambda^l_{k'k}\;\mbox{\boldmath$\sigma$}_j \times \mbox{\boldmath$\sigma$}_k\cdot{\bf B}.
\end{eqnarray} 
In the last two equations, $Q_l$ is the charge of meson $l$, and $J_l$ is the integral
\begin{equation}
J_l = \frac{\beta^2}{16\pi^2f^2}\int_0^\infty dk\,\frac{k^4}{(k^2+M_l^2)^2} F^2(k^2).
\end{equation}

The flavor factor in Eq.\ (\ref{10b}) is equivalent to that in Eq.\ (\ref{lambda_M_lambda}) since $QM=-M/3$. The factor in Eq.\ (\ref{10d}) is given in Eq.\ (\ref{lambda_lambda_sum}). That in Eq. (\ref{10f}) follows from the result in Eq.\ (\ref{I_lambda'_lambda}) by multiplication by $Q$, with the result
\begin{eqnarray}
\sum_l I'_l\,\left(Q\lambda^l\right)_{j'j}\lambda^l_{k'k} &=& 2I_\pi\,Q_{j'k}\openone_{k'j}  - \left(I_\pi-\frac{1}{3}I_\eta\right)\,Q_{j'j}\openone_{k'k}  \nonumber \\
&& - 2\left(I_\pi-I_K\right) \left(Q_{j'k}M_{k'j} + \left(QM\right)_{j'k}\openone_{k'j}\right)   \nonumber \\ 
&& +\left(I_\pi-I_\eta\right) \left(Q_{j'j}M_{k'k} + \left(QM\right)_{j'j}\openone_{k'k}\right) \\
&& +\left(I_\pi-4I_K+3I_\eta\right)\left(QM\right)_{j'j}M_{k'k}.  \nonumber
\end{eqnarray}
The remaining flavor factors are new. We find after some calculation that
\begin{eqnarray}
\label{lambda_Q_lambda}
\sum_lI'_l\,\left(\lambda^l Q\lambda^l\right)_{k'k} &=& \frac{2}{3}\left(I'_\pi-I'_K\right)\openone_{k'k} - \frac{1}{3}\left(3I'_\pi - 4I'_K+I'_\eta\right)M_{k'k} - \frac{1}{3}\left(3I'_\pi-I'_\eta\right)Q_{k'k},
\\
\label{JQ_lambda_lambda}
\sum_l Q_lJ_l\,\left(\lambda^l \lambda^l\right)_{k'k} &=& -\frac{2}{3}\left(J_\pi-J_K\right)\openone_{k'k} + \left(4J_\pi+2J_K\right)Q_{k'k} + 2\left(J_\pi-J_K\right)M_{k'k},
\\
\label{JQ_lambda'_lambda}
\sum_l Q_lJ_l\; \lambda^l_{j'j}\lambda^l_{k'k} &=& J_\pi\left(\lambda^2_{j'j} \lambda^1_{k'k}-\lambda^1_{j'j} \lambda^2_{k'k}\right) + J_K\left( \lambda^4_{j'j} \lambda^3_{k'k} - \lambda^3_{j'j} \lambda^4_{k'k}\right). \end{eqnarray}

Using these results, we find that the one-body graphs in Figs.\ \ref{fig12}\,(a), (b) and the corresponding renormalization graphs involve only the one-body flavor structures $\openone_{k'k}$, $Q_{k'k}$,  $(MQ)_{k'k}$, and $M_{k'k}=-3(MQ)_{k'k}$, all multiplied by $\mbox{\boldmath$\sigma$}_k \!\cdot\! {\bf B}$. The operators $Q_k\,\mbox{\boldmath$\sigma$}_k \!\cdot\! {\bf B}$ and $(MQ)_k\,\mbox{\boldmath$\sigma$}_k \!\cdot\! {\bf B}$ already appear as invariants in Fig.\ \ref{fig10}, and define the effective quark moment operator $\mbox{\boldmath$\mu$}_k=(\mu_1Q+\mu_2QM)_k \mbox{\boldmath$\sigma$}_k$. The corresponding components of the loop corrections alter the values of $\mu_1$ and $\mu_2$, but add no new structure. 

In contrast, the one-body operator $\openone_{k'k}\,\mbox{\boldmath$\sigma$}_k \!\cdot\! {\bf B}$ does not appear explicitly in Fig.\ \ref{fig10}. The sum of those terms over the three quark lines gives an effective baryon-level moment operator $(\mbox{\boldmath$\sigma$}_i +\mbox{\boldmath$\sigma$}_j +\mbox{\boldmath$\sigma$}_k)\!\cdot\!{\bf B} = \mbox{\boldmath$\sigma$}\!\cdot\! {\bf B}$ where $\mbox{\boldmath$\sigma$}$ is the total spin operator. This operator gives equal contributions to all the diagonal octet moments, and no contribution to the $\Sigma^0\Lambda$ transition moment. While $\mbox{\boldmath$\sigma$}\!\cdot\!{\bf B}$ does not appear in our earlier list, the modified octet moments continue to satisfy the Okubo sum rule in Eq.\ (\ref{Okubo_sum}), so should be expressible in terms of the existing structures. The necessary transformation is given by the inverse of the upper left $7\times 7$ matrix in Table \ref{table:octet_moments}. There is no new structure.\footnote{Specifically, the contribution of the invariant $\mbox{\boldmath$\sigma$}\!\cdot\!{\bf B}$ to the octet moments is the same as that of the combination
\begin{displaymath}
\left(-\frac{3}{2}{\bf m_1} + \frac{5}{2}{\bf m_3} +\frac{3}{2}{\bf m_4} + \frac{9}{4}{\bf m_5} - 3{\bf m_6} - \frac{7}{4}{\bf m_7}\right)\!\cdot\!{\bf B}
\end{displaymath}
of the operators in Eqs.\ (\ref{basic})-(\ref{QN_s}).}

We can treat the loop corrections to the two-body Thomas term shown in Figs.\ \ref{fig12}\,(c) and (d) similarly. The graph in Fig.\ \ref{fig12}\,(c) involves the structures $Q_{j'j}\openone_{k'k}$, $M_{j'j}\openone_{k'k}=-3(QM)_{j'j}\openone_{k'k}$, and $\openone_{j'j} \openone_{k'k}$, all multiplied by $\mbox{\boldmath$\sigma$}_k \!\cdot\! {\bf B}$. The first two simply modify the parameters associated with the diagrams in Figs.\ \ref{fig10}\,(c) and (f). The third simply modifies the coefficient of $\mbox{\boldmath$\sigma$}\!\cdot\! {\bf B}$, so is reducible as above to existing structures.

In the case of the diagram in Fig.\ \ref{fig12}\,(d), we encounter the structures $Q_{j'j}\openone_{k'k}$ and  $Q_{j'j}M_{k'k}$ multiplied by $\mbox{\boldmath$\sigma$}_k \!\cdot\! {\bf B}$. The first modifies the Thomas term in Fig.\ \ref{fig10}\,(c), and the second, the term in Fig.\ \ref{fig10}\,(e). There are no new contributions.

The meson-exchange diagrams in Fig.\ \ref{fig12}\,(f) are more interesting. Combining the contributions for the two time orderings and the corresponding wave function renormalization terms, we obtain a total contribution 
\begin{equation}
\label{10f_term}
\frac{1}{3}\sum_lI'_l\left[Q,\,\lambda^l\right]_{j'j}\lambda^l_{k'k} \,\left[\left(\mbox{\boldmath$\sigma$}_k\!\cdot\!{\bf B}\right) \left(\mbox{\boldmath$\sigma$}_j \!\cdot\! \mbox{\boldmath$\sigma$}_k\right) - \left(\mbox{\boldmath$\sigma$}_j \!\cdot\! \mbox{\boldmath$\sigma$}_k\right) \left(\mbox{\boldmath$\sigma$}_k\!\cdot\!{\bf B}\right)\right].
\end{equation}
The spin operator in this expression could be reduced to $2i\mbox{\boldmath$\sigma$}_j\times\mbox{\boldmath$\sigma$}_k\!\cdot\!{\bf B}$, a form which makes it clear that operator connects singlet and triplet spin states of quarks $j$ and $k$. However, it is more convenient to retain the spin structure in Eq.\ (\ref{10f_term}) because this allows us to use the exchange-operator method given following Eq.\ (\ref{P_2}) to eliminate the mixed indices in the flavor factor. Combining the results with those for the graphs with the topology of Fig.\ \ref{fig12}\,(f) with the labels $j$ and $k$ interchanged, we obtain a total contribution
\begin{eqnarray}
\frac{8}{3}\left\{\,I'_\pi\left(Q_{j'j}\openone_{k'k} - \openone_{j'j}Q_{k'k} \right) - \left(I'_\pi-I'_K\right)\left[(MQ)_{j'j}\openone_{k'k} - \openone_{j'j}(MQ)_{k'k}\right]\right. \nonumber 
\\
- \left.\left(I'_\pi-I'_K\right)\left(Q_{j'j}M_{k'k}-M_{j'j}Q_{k'k}\right) \right\}\left( \mbox{\boldmath$\sigma$}_k - \mbox{\boldmath$\sigma$}_j\right)\cdot{\bf B}. 
\end{eqnarray}
The singlet-triplet character of the spin factor is again clear. The first term, when multiplied out, can be expressed as a sum of the two-body structures in Figs.\ \ref{fig10}\,(a) and (c), and the second and third terms, as sums of the structures in Figs.\ \ref{fig10}\,(b), (f), and (d), (e), respectively. The meson exchange therefore contributes no new structure in the general case.

We turn next to the diagrams in Figs.\ \ref{fig13} which involve direct meson-photon interactions. The magnetic-moment contributions from Fig.\ \ref{fig13}\,(a), which we obtain using Eqs.\ ({11a}) and (\ref{JQ_lambda_lambda}), have a one-body structure expressed in terms of factors $\openone$, $Q$, and $M$. These modify the preceding results, but do not contribute anything intrinsically new.

The meson exchange-current diagrams in Fig.\ \ref{fig13}\,(b) are more subtle. There is now a nontrivial $\lambda$ matrix at each vertex corresponding to a switch of the associated quark charge by $\pm 1$, a two-body structure,  and a total contribution
\begin{equation}
\label{meson_exchange_current}
\frac{1}{3}\left[J_\pi\left(\lambda^2_{j'j} \lambda^1_{k'k}-\lambda^1_{j'j} \lambda^2_{k'k}\right) + J_K\left( \lambda^4_{j'j} \lambda^3_{k'k} - \lambda^3_{j'j} \lambda^4_{k'k}\right)\right]\,\mbox{\boldmath$\sigma$}_j \times \mbox{\boldmath$\sigma$}_k\cdot{\bf B}
\end{equation}
to the baryon moment operator. If we write out the $\lambda$ matrices explicitly, we find that
\begin{eqnarray}
\lambda^2_{j'j} \lambda^1_{k'k}-\lambda^1_{j'j} \lambda^2_{k'k} &=& 2i\left( \delta_{j'2}\delta_{k2}\,\delta_{k'1}\delta_{j1} - \delta_{j'1}\delta_{k1}\,\delta_{k'2}\delta_{j2} \right) \nonumber \\
&=& \frac{2}{3}i\Bigl[\openone_{j'k}\left(Q+M\right)_{k'j}-\left(Q+M\right)_{j'k} \openone_{k'j}  \nonumber \\ 
&& +\left(Q_{j'k}M_{k'j} - M_{j'k}Q_{k'j}\right)\Bigr], 
 \\
\lambda^4_{j'j} \lambda^3_{k'k} - \lambda^3_{j'j} \lambda^4_{k'k} &=& 2i\left(\delta_{j'3}\delta_{k3}\,\delta_{k'1}\delta_{j1} - \delta_{j'1}\delta_{k1}\,\delta_{k'3}\delta_{j3}\right) \nonumber \\
&=& \frac{2}{3}i\left[M_{j'k}\left(\openone + Q\right)_{k'j} - \left(\openone + Q\right)_{j'k}M_{k'j}\right],
\end{eqnarray}
where we have used the identities
\begin{equation}
\delta_{j'1}\delta_{k1}= \frac{1}{3}\left(\openone+Q\right)_{j'k}, \quad \delta_{j'2}\delta_{k2} = \frac{1}{3}\left(2\openone - Q -3M\right)_{j'k}, \quad \delta_{j'3}\delta_{k3}=M_{j'k}
\end{equation}
to combine factors. Finally, using the permutation operator in Eq.\ (\ref{P_2}) and the relation
\begin{equation}
\frac{1}{2}\left(1 + \mbox{\boldmath$\sigma$}_j \!\cdot\! \mbox{\boldmath$\sigma$}_k \right) \mbox{\boldmath$\sigma$}_j\times\mbox{\boldmath$\sigma$}_k\cdot{\bf B} = -i\left(\mbox{\boldmath$\sigma$}_j-\mbox{\boldmath$\sigma$}_k\right)\!\cdot\! {\bf B},
\end{equation}
we find that the exchange-current contributions from Fig. \ref{fig13}\,(b) can be written as
\begin{eqnarray}
\frac{2}{9}\left\{J_\pi\left[\left(Q+M\right)_{j'j}\openone_{k'k} - \openone_{j'j}\left(Q+M\right)_{k'k} + M_{j'j}Q_{k'k} - Q_{j'j}M_{k'k}\right]\right. \nonumber \\
\left. +J_K\left[\openone_{j'j}M_{k'k} -M_{j'j}\openone_{k'k}+Q_{j'j}M_{k'k} - M_{j'j}Q_{k'k}\right]\right\} \left(\mbox{\boldmath$\sigma$}_j - \mbox{\boldmath$\sigma$}_k\right)\!\cdot\! {\bf B}.
\end{eqnarray}
This is completely expressible in terms of the operators in Fig. \ref{fig10}, and despite the appearance of the $\lambda$ matrices, nothing new is added.
 
The situation changes when we consider loop corrections to the smaller terms corresponding to the diagrams in Figs.\ \ref{fig10}\,(d)-(f) and \ref{fig11}. We now obtain terms involving two factors of $M$ on different quark lines. Such terms arise, for example, from loop corrections to the charge vertex in Figs.\ \ref{fig10}\,(d) and (e). The quadratic $MM$ terms are not included among the O($M$) invariants in Eqs.\ (\ref{one-body_moments}) or (\ref{chiral_terms}), and cannot be reduced to linear combinations of those structures. We have not made a detailed calculation of these corrections, but find several possible structures including
\begin{eqnarray}
&&\left(M_i+M_j\right)M_k\mbox{\boldmath$\sigma$}_k + {\rm permutations} = \left(N_s-1\right)\sum_lM_l\mbox{\boldmath$\sigma$}_l, \nonumber \\
&& M_iM_j\mbox{\boldmath$\sigma$}_k + {\rm permutations} = \frac{1}{2}N_s\left(N_s-1\right)\mbox{\boldmath$\sigma$} - \left(N_s-1\right)\sum_lM_l\mbox{\boldmath$\sigma$}_l, \\
&&Q_iM_jM_k\left(\mbox{\boldmath$\sigma$}_j+\mbox{\boldmath$\sigma$}_k\right)
+ {\rm permutations} = \left(Q-2Q_s\right)\left(N_s-1\right) \sum_lM_l\mbox{\boldmath$\sigma$}_l, \nonumber
\end{eqnarray}
and similar terms with extra factors of $\mbox{\boldmath$\sigma$}' \!\cdot \mbox{\boldmath$\sigma$}''$ connecting the $M$ vertices. These contribute only to the $\Xi^0$ and $\Xi^-$ moments when acting on octet states. It follows from the coefficients in the first seven columns  in Table \ref{table:octet_moments} that a term which contributes only to the $\Xi^-$ moment can be expressed in terms of the combination $3{\bf m}_5 + {\bf m}_7$ of the operators in Eqs.\ (\ref{Q*Msigma}) and (\ref{QN_s}). Only the change in the $\Xi^0$ moment corresponds to a new invariant. This can be written for the octet as the effective operator
\begin{equation}
\label{m_8}
{\bf m}_8=(Q+1)N_s(N_s-1) \mbox{\boldmath$\sigma$}/2, 
\end{equation}
corresponding to the single entry in the last column of Table \ref{table:octet_moments}.

We conclude this section by reemphasizing that none of the diagrams in Figs. \ref{fig12} and \ref{fig13} introduces any spin/flavor structures beyond those already included in the one- and two-body contributions to the moments in Fiq. \ref{fig10}. New structures quadratic in $M$ only appear when we consider loop corrections to the two- or three-body terms that are already of order $M$. The meson-current terms are particularly interesting in this respect. These are formally of order $m_s^{1/2}$ in the chiral expansion and are supposed to give the leading corrections to the moments in ChPT \cite{Caldi,Jen-moments,Mei-moments}. However, when we calculate as here, starting with the symmetrical limit of degenerate octet and decuplet baryons with SU(6) couplings, we find that these terms can be absorbed completely in redefinitions of the short-distance parameters associated with the diagrams in Fig.\ \ref{fig10}, and contribute nothing that could not be described already at tree level. We find this quite striking. 

\addtocounter{footnote}{-2}

The situation is different if one starts from a model with a reduced set of input parameters, for example, the Coleman-Glashow model \cite{Caldi,Jen-moments,Mei-moments} or the quark model \cite{DH-loop-moments} which initially involve only the diagrams in Figs.\ \ref{fig10}\,(a) and (c), and (a) and (b), respectively. The loop corrections then generate complete set of invariants discussed in connection with Figs.\ \ref{fig12} and \ref{fig13} as dynamical output. The success of the model in fitting the data then determines whether or not further contributions must be included from the possible short-distance parameters.\footnotemark 

The same general conclusions apply to the baryon axial currents. These have the same formal structure as the magnetic moments.
\addtocounter{footnote}{+1}
\section{CONCLUSIONS}
\label{sec:conclusions}

Our intent here was to develop methods which can be used fairly easily to analyze the structure and significance of meson loop corrections in the context of HBChPT. We showed, in particular, that the major part of the loop corrections to the baryon masses and magnetic moments can be absorbed in adjustments of the short-distance chiral parameters in the expansions of these quantities to O($m_s$). The only quantities that need to be calculated are those of O($M^2$) or higher in the quark mass matrix $M$. These can be extracted rather simply. It is unfortunate that such methods were not developed earlier. An enormous amount of work has gone into general calculations of loop corrections to the baryon masses \cite{Jen-HBChPT,Jen-masses,Bernard-masses,Bor-Mei-masses,Don-masses,%
DH-masses,phuoc-diss,jaczko-diss} and moments \cite{Caldi,Gasser,Krause,%
Jen-moments,Luty,Daietal,Mei-moments,DH-ChPTmoments,DH-loop-moments,%
Ha-moments} in ChPT and in $1/N_c$ expansions \cite{Luty,Daietal,Dashen,JenkinsManohar}, but to rather little effect when used with the general O($m_s$) parametrization. The key problem in obtaining a useful, predictive expansion is the determination of at least the some of the chiral parameters, as, for example, in dynamical models such as those in \cite{phuoc-diss,DHJ}, in $1/N_c$ expansions \cite{Luty,Daietal,Dashen,JenkinsManohar}, or otherwise. 

We have treated the baryon octet and decuplets as degenerate in our calculations, with the SU(6)-symmetric meson-baryon couplings that result from the weakness of the spin-spin forces between quarks \cite{DHJ}, and have expanded to first order in the octet-decuplet mass difference along with the other mass parameters. This approach also contributes to the relative simplicity of our results. The apparently small deviations of the couplings from their SU(6) limits can be handled by similar methods using projection operators, as noted earlier and discussed in \cite{DHJ}.  We remark that there are also O($M$) corrections to the axial couplings that we and other authors have not considered. 

We note, finally, that the momentum structure of the residual loop corrections should be examined in any calculation. As shown for the baryon masses in Sec.\ \ref{subsec:momentum_structure}, the supposedly low-momentum loop integrals can contain rather large high-momentum or short-distance components for point baryons. The use of an appropriate form factor to incorporate the extended structure of the baryons eliminates this problem.

\acknowledgments
This work was supported in part by the U.S. Department of Energy under Grant No.\ DE-FG02-95ER40896, and in part by the University of Wisconsin Graduate School with funds granted by the Wisconsin Alumni Reseach Foundation. One of the authors (LD) would like to thank the Aspen Center for Physics for its hospitality while parts of this work were done.

%\end{document}

% FIGURES

\newpage

\input epsf

% Fig. 1 %

% Fig. 2 %

% Fig. 3 %

% Fig. 4 %

% Fig. 5 %

% Fig. 6 %

% Fig. 7 %

% Fig. 8 %

% Fig. 9 %

% Fig. 10 %

% Fig. 11 %

% Fig. 12 %

% Fig. 13 %

\newpage

\begin{table}
\caption{The coefficients for the expression of the octet baryon moments in terms of the operators in Eqs.\ (\protect\ref{basic})-(\protect\ref{QN_s}) and \protect(\ref{m_8}). The moment of the $\Sigma^0$ is determined by isospin invariance, $\mu_{\Sigma^0}=\frac{1}{2}(\mu_{\Sigma^+}+\mu_{\Sigma^-})$.}
\begin{tabular}{ccccccccc}
Baryon & $\mu_1$ & $\mu_2$ & $\mu_3$ & $\mu_4$ & $\mu_5$ & $\mu_6$ & $\mu_7$ & $\mu_8$ \\
\hline
$p$ & 1  & 0  & 1  & 0  & 0  &  0 & 0 & 0\\
$n$ & -2/3  & 0  & 0  & 0  & 0  & 0  & 0 & 0 \\
$\Sigma^+$ & 1 & 1/9  & 1  & 1  & -1/3  & -1/3  & 1 & 0 \\
$\Sigma^-$ & -1/3  & 1/9  & -1  & -1/3  & 1/3  & -1/3  & -1 & 0 \\
$\Xi^0$ & -2/3  & -4/9 & 0  & -4/3  & 0  & -2/3  & 0 & 1 \\
$\Xi^-$ & -1/3  &  -4/9  & -1  & -2/3  & -2/3  & -2/3  & -2 & 0 \\
$\Lambda$ & -1/3  & -1/3  & 0  & -1/3  & 0 & -1/3  & 0 & 0 \\
$\Sigma^0\Lambda$ & $1/\sqrt{3}$  & 0  & 0  & $1/\sqrt{3}$  &  0 &  0 & 0 & 0 \\
\end{tabular}
\label{table:octet_moments}
\end{table}

\end{document}